\documentclass[fleqn,usenatbib]{mnras}
\usepackage{newtxtext,newtxmath}
\usepackage[T1]{fontenc}
\usepackage{ae,aecompl}

\usepackage{graphicx}
\usepackage{amsmath}
\usepackage{amssymb}
\usepackage{bm}
\usepackage{upgreek}
\usepackage{subfig}
\usepackage{relsize}

\newcommand{\ph}[1]{\phantom{#1}}

\newcommand{\sourceXLIV}[0]{APMUKS(BJ) B001307.84\allowbreak{}$-$384221.2}
\newcommand{\sourceCLXXVII}[0]{APMUKS(BJ) B235959.18\allowbreak{}$-$220253.5}
\newcommand{\sourceCLXXXV}[0]{2MASX J00140036\allowbreak{}$-$1749526}
\newcommand{\sourceXCIX}[0]{2dFGRS S491Z111}
\newcommand{\sourceSV}[0]{GALEXASC J005501.01\allowbreak{}$-$231008.9}

\renewcommand{\ion}[2]{\ensuremath{\text{#1\hspace{0.15em}{\smaller #2}}}}
\renewcommand{\vec}[1]{\ensuremath{\bm{#1}}}

\title[A deep \ion{H}{I} survey of the Sculptor group]{A deep Parkes \ion{H}{I} survey of the Sculptor group and filament: \ion{H}{I} mass function and environment}
\author[T.~Westmeier et al.]{T. Westmeier$^{1}$\thanks{E-mail:
tobias.westmeier@uwa.edu.au}, D.~Obreschkow$^{1}$, M.~Calabretta$^{2}$, R.~Jurek$^{3}$, B.~S.~Koribalski$^{2}$, \newauthor M.~Meyer$^{1,4}$, A.~Musaeva$^{5}$, A.~Popping$^{1,4}$, L.~Staveley-Smith$^{1,4}$, O.~I.~Wong$^{1,4}$ \newauthor and A.~Wright$^{1,6}$\\
$^{1}$International Centre for Radio Astronomy Research (ICRAR), The University of Western Australia, 35 Stirling Highway,\\Crawley WA 6009, Australia\\
$^{2}$ATNF, CSIRO Astronomy and Space Science, PO Box 76, Epping NSW 1710, Australia\\
$^{3}$Department of Physics, School of Mathematics and Physics, The University of Queensland, Brisbane QLD 4072, Australia\\
$^{4}$ARC Centre of Excellence for All-sky Astrophysics (CAASTRO)\\
$^{5}$Sydney Institute for Astronomy, School of Physics A28, The University of Sydney NSW 2006, Australia\\
$^{6}$Argelander-Institut f\"{u}r Astronomie, Rheinische Friedrich-Wilhelms-Universit\"{a}t Bonn, Auf dem H\"{u}gel 71, 53121 Bonn, Germany}
\begin{document}

\date{Accepted 2017 September 1. Received 2017 August 31; in original form 2017 June 19}

\pagerange{\pageref{firstpage}--\pageref{lastpage}} \pubyear{2017}

\maketitle

\label{firstpage}

\begin{abstract}
  We present the results of a deep survey of the nearby Sculptor group and the associated Sculptor filament taken with the Parkes 64-m radio telescope in the 21-cm emission line of neutral hydrogen. We detect 31~\ion{H}{I} sources in the Sculptor group/filament, eight of which are new \ion{H}{I} detections. We derive a slope of the \ion{H}{I} mass function along the Sculptor filament of $\alpha = -1.10^{+0.20}_{-0.11}$, which is significantly flatter than the global mass function and consistent with the flat slopes previously found in other low-density group environments. Some physical process, such as star formation, photoionisation or ram-pressure stripping, must therefore be responsible for removing neutral gas predominantly from low-mass galaxies. All of our \ion{H}{I} detections have a confirmed or tentative optical counterpart and are likely associated with luminous rather than `dark' galaxies. Despite a column density sensitivity of about $4 \times 10^{17}~\mathrm{cm}^{-2}$, we do not find any traces of extragalactic gas or tidal streams, suggesting that the Sculptor filament is, at the current time, a relatively quiescent environment that has not seen any recent major interactions or mergers.
\end{abstract}


\section{Introduction}

In the currently favoured Lambda Cold Dark Matter ($\Lambda$CDM) paradigm, galaxies do not evolve in isolation, but instead form along cosmic filaments of dark matter \citep{Efstathiou1990,Suginohara1991,Gnedin1996}. The environment in which galaxies form and evolve is thought to have a strong impact on their most fundamental properties, including their gas accretion (e.g., \citealt{vandeVoort2017}) and star formation (e.g., \citealt{Rettura2010,Guglielmo2015}) histories. As star formation is fuelled by gas, probing the neutral gas content of galaxies in different environments -- ranging from voids to the densest clusters -- is a logical way of assessing the effectiveness of gas removal in environments of different density \citep{Kawata2008,Fumagalli2008,Jaffe2015} and its contribution to the quenching of star formation \citep{Yozin2015}.

\begin{table*}
  \caption{Parameters of the \ion{H}{I} data sets used in this work. The columns denote: FWHM of the gridded beam ($\theta$), velocity resolution for \ion{H}{I} emission at redshift zero ($\Delta v$), RMS noise level at the original velocity resolution ($\sigma_{\rm RMS}$), $5 \sigma$ \ion{H}{I} mass sensitivity for unresolved sources at a distance of $1~\mathrm{Mpc}$ ($M_{\ion{H}{I}}$), and $5 \sigma$ \ion{H}{I} column density sensitivity for emission filling the beam ($N_{\ion{H}{I}}$). A line width of $26.4~\mathrm{km \, s}^{-1}$ was assumed for all \ion{H}{I} mass and column density calculations.}
  \label{tab_data}
  \begin{tabular}{lrrrrrr}
    \hline
    Data set & $\theta$ & $\Delta v$                & $\sigma_{\rm RMS}$ & $\sigma_{\rm RMS}$ & $M_{\ion{H}{I}}$     & $N_{\ion{H}{I}}$     \\
             & (arcmin) & ($\mathrm{km \, s}^{-1}$) & (mJy)              & (mK)               & ($\mathrm{M}_{\sun}$)         & ($\mathrm{cm}^{-2}$) \\
    \hline
    Parkes   &   $15.5$ &                     $1.6$ &             $11.0$ &              $7.7$ &  $8.4 \times 10^{4}$ & $4.6 \times 10^{17}$ \\
    HIPASS~2 &   $15.5$ &                    $26.4$ &              $4.0$ &              $2.8$ & $12.5 \times 10^{4}$ & $6.7 \times 10^{17}$ \\
    Combined &   $15.5$ &                    $26.4$ &              $2.5$ &              $1.8$ &  $7.8 \times 10^{4}$ & $4.3 \times 10^{17}$ \\
    \hline
  \end{tabular}
\end{table*}

A promising method previously used to measure the impact of the environment on the neutral gas content of galaxies is to study the variations in the slope and turnover point of the \ion{H}{I} mass function (HIMF) for environments of different density \citep{Rosenberg2002}. Initial attempts by \citet{Zwaan2005} and \citet{Springob2005} yielded conflicting results. From their analysis based on 4315~galaxies detected in the \ion{H}{I} Parkes All-Sky Survey (HIPASS; \citealt{Barnes2001,Meyer2004}), \citet{Zwaan2005} found tentative evidence for a steepening of the slope of the HIMF in high-density environments. \citet{Springob2005}, however, found the opposite effect of a steeper slope in low-density environments based on a specially selected sample of 2771~galaxies. The tension between these two results may in part be due to the different methodologies used in defining environmental density. The density definition used by \citet{Zwaan2005} was based on the HIPASS catalogue itself, while \citet{Springob2005} utilised the IRAS Point Source Catalog Redshift (PSCz) catalogue by \citet{Saunders2000} to define density. \ion{H}{I} catalogues in particular may not be a suitable tracer of environmental density due to the \ion{H}{I} deficiency of galaxies in high-density environments \citep{Solanes2001}.

More recently, a study by \citet{Jones2016} based on the 70~per cent catalogue of the Arecibo Legacy Fast ALFA (ALFALFA) survey \citep{Giovanelli2005} found no evidence of any change in the slope of the HIMF with environment, although a small environmental effect on the turnover mass, $M_{\ion{H}{I}}^{\star}$, was observed. They also caution against comparing their result with that of \citet{Zwaan2005} due to the different definitions of environmental density used by the two studies. A similar study by \citet{Moorman2014} based on ALFALFA data also finds a small environmental effect on the turnover mass of the HIMF as well as a slightly steeper slope for their high-density sample. The aforementioned results from global studies of the HIMF are generally in disagreement with studies of individual galaxy groups which usually find a flat lower end of the HIMF ($\alpha \approx -1$, e.g. \citealt{Freeland2009,Kilborn2009,Pisano2011}). An exception is the Leo~I group with a rather steep slope of $\alpha = -1.41_{-0.1}^{+0.2}$ \citep{Stierwalt2009}.

Here, we present the results of a deep \ion{H}{I} study of the Sculptor group region with the aim to obtain a census of galaxies down to an \ion{H}{I} mass sensitivity of the order of $10^{7}~\mathrm{M}_{\sun}$ and compare the slope of the resulting HIMF with previous studies, both of individual groups as well as global galaxy samples. Due to its proximity, the Sculptor group is particularly suitable for such a study, as we are able to achieve an \ion{H}{I} mass sensitivity superior to most other group studies. At the same time, the Sculptor group can be studied with a large, sensitive single-dish telescope without running into issues of source confusion or missing flux on large angular scales.

A peculiarity of the Sculptor group is the fact that it does not constitute a single, compact, gravitationally bound group, but rather a collection of separate sub-groups along a coherent filament \citep{Arp1985,Jerjen1998} that shall be referred to as the `Sculptor filament' throughout this paper. In fact, the Local Group forms a part of this Sculptor filament \citep{Arp1985,Karachentsev2003}, with the nearest Sculptor sub-group, consisting of NGC~55, NGC~300 and a few dwarf galaxies, being only about $2~\mathrm{Mpc}$ away from us. As a consequence of our location within the Sculptor filament, we essentially see the sub-groups of the filament lined up along the line of sight like pearls on a string. Due to this special geometry, there is a significant variation in distance along the Sculptor filament, with the farthest members considered here having a distance of about $15~\mathrm{Mpc}$ from the Milky Way. This stretch in distance leads to a significant change in spatial resolution (factor of $\approx 7$) and \ion{H}{I} mass sensitivity (factor of $\approx 50$) between the near and far ends of the Sculptor filament.

The \ion{H}{I}~emission of all major Sculptor group galaxies has been studied in detail with the VLA in a series of papers by \citet{Carignan1990a,Carignan1990b}, \citet{Puche1990,Puche1991a,Puche1991b} and \citet{Chemin2006}. More sensitive \ion{H}{I}~observations of NGC~300 and NGC~55 with the Australia Telescope Compact Array (ATCA) were presented by \citet{Westmeier2011,Westmeier2013}, and \ion{H}{I}~detections of several dwarf galaxies across the Sculptor group were reported by \citet{Cote1997} and \citet{Bouchard2005} based on observations with the Parkes radio telescope. Additional ATCA \ion{H}{I} observations of many Sculptor group galaxies that were taken as part of the Local Volume \ion{H}{I} Survey (LVHIS) will be presented by Koribalski et al. (in prep.). While these observations have been focussing on individual galaxies, the survey presented here constitutes the first ever deep \ion{H}{I} survey of the entire Sculptor group region, covering an area of over $2~\mathrm{h}$ in right ascension and almost $30~\mathrm{deg}$ in declination.

This paper is organised as follows. Our observations and data reduction strategy are outlined in Section~\ref{sect_observations}. In Section~\ref{sect_analysis} we describe our data analysis procedure before presenting our main results and the HIMF of the Sculptor filament in Section~\ref{sect_results}. In Section~\ref{sect_discussion} we discuss our observational results, before summarising our main findings and conclusions again in Section~\ref{sect_summary}.

\begin{figure*}
  \includegraphics[width=0.68\linewidth]{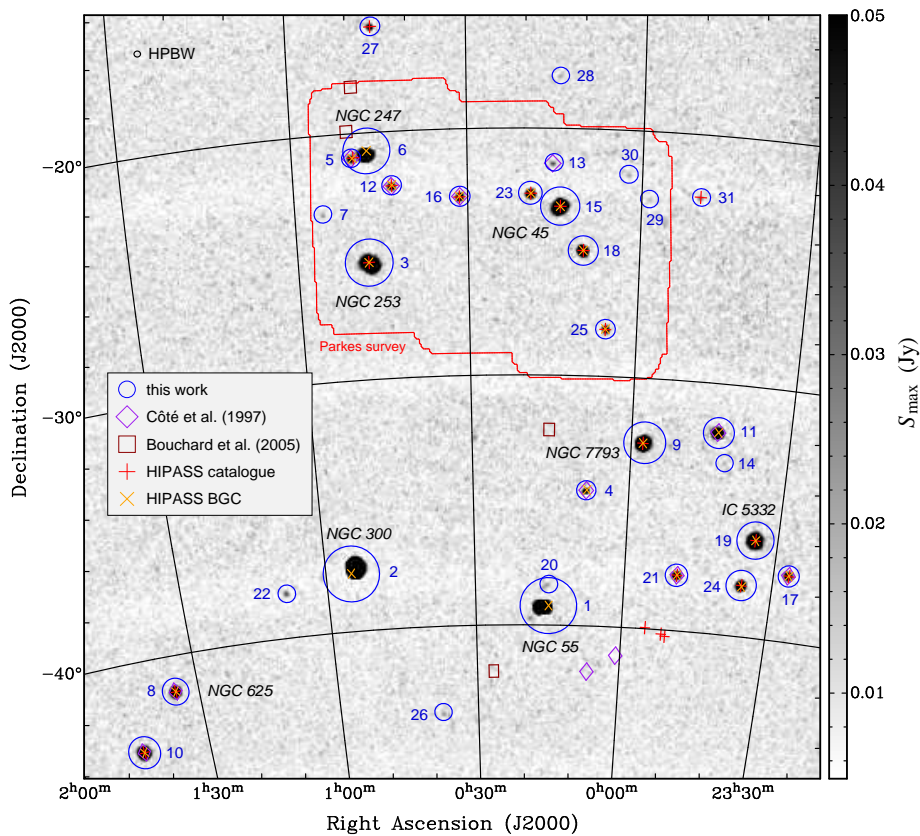}
  \caption{Peak flux density map of the Sculptor group in the velocity range of $220~\mathrm{km \, s}^{-1} < \mathrm{c}z < 1200~\mathrm{km \, s}^{-1}$ based on the HIPASS~2 data. All detected \ion{H}{I}~sources that were identified as extragalactic are marked with blue circles, the radius of which is proportional to the logarithm of their integrated flux, and blue numbers corresponding to the source~ID in Table~\ref{table_hipass_source_list}. Note that some flux is missing from nearby galaxies, in particular NGC~55, 247 and 300, as their emission extends to velocities of $\mathrm{c}z < 220~\mathrm{km \, s}^{-1}$. The red outline marks the region covered by our deep follow-up survey with the Parkes telescope (see Fig.~\ref{fig_smax2}). Sources listed in the main HIPASS catalogue \citep{Meyer2004} and the HIPASS Bright Galaxy Catalog \citep{Koribalski2004} are marked with red and orange crosses, respectively. \ion{H}{I}~detections from \citet{Cote1997} and \citet{Bouchard2005} are marked with purple diamonds and brown squares, respectively.}
  \label{fig_smax}
\end{figure*}

\section{Observations and data reduction}
\label{sect_observations}

Deep \ion{H}{I}~observations of the northern Sculptor group region were carried out between May and September 2013 using the 64-m Parkes radio telescope. The survey area of approximately $15 \times 10$~degrees was chosen to cover the galaxies of the northern Sculptor group around NGC~247/253 and NGC~45. For practical reasons, the area was split up into six separate sub-regions, each covering about $5 \times 5$~degrees. Using the 20-cm multi-beam receiver, each sub-region was scanned twelve times in both right ascension and declination at a rate of 34~arcsec per second, with the thirteen feed horns of the receiver rotated on the sky by 15~degrees. The individual scan lines in the final, combined map are separated by $5~\mathrm{arcmin} \approx \mathrm{HPBW} / 3$, thus forming a dense basket-weaving pattern to minimise scanning artefacts. The total integration time per pointing amounts to just over $1~\mathrm{h}$, resulting in an RMS noise level of about $11~\mathrm{mJy}$ (equivalent to $7.7~\mathrm{mK}$) at the original frequency resolution of $7.8~\mathrm{kHz}$ (equivalent to $1.6~\mathrm{km \, s}^{-1}$ for \ion{H}{I} emission at redshift zero).

The data were reduced using the standard Parkes data reduction pipeline, \textsc{Livedata}/\textsc{Gridzilla} \citep{Barnes2001}. In order to calibrate the bandpass, we used \textsc{Livedata}'s median estimator in combination with the extended source method. This would create an emission-free reference bandpass per scan line that was used for bandpass correction of each individual integration. After successful bandpass calibration, the data were gridded into a single data cube using the weighted median estimator in \textsc{Gridzilla}. The resulting data cube has a beam size of $15.5~\mathrm{arcmin}$ and a spatial pixel size of $4~\mathrm{arcmin}$, while the spectral axis describes frequency in the barycentric rest frame, covering a total of $8~\mathrm{MHz}$ across a velocity range of ${-300} \lesssim \mathrm{c}z \lesssim 1300~\mathrm{km \, s}^{-1}$. In a final step, a polynomial of degree~5 was fitted to the line-free channels of every spectrum in the cube to remove any remaining bandpass variations.

In addition we made use of data from HIPASS~2, an improved version of the \ion{H}{I} Parkes All-Sky Survey \citep{Barnes2001,Calabretta2014}. In the region of the Sculptor group, HIPASS~2 incorporates additional data which significantly reduces the RMS noise level compared to the original HIPASS data. The original spectral channel width of HIPASS~2 is $62.5~\mathrm{kHz}$ at a spectral resolution of $75.6~\mathrm{kHz}$ (corresponding to $13.2$ and $16.0~\mathrm{km \, s^{-1}}$, respectively, for \ion{H}{I} emission at redshift zero). In order to alleviate the effect of Gibbs ringing near strong Galactic \ion{H}{I} emission, we applied a Hann filter along the spectral axis of the HIPASS~2 data, yielding a final RMS level of about $4.0~\mathrm{mJy}$ at a velocity resolution of $26.4~\mathrm{km \, s^{-1}}$. A peak flux density map of the HIPASS~2 data cube is displayed in Fig.~\ref{fig_smax}, which also includes an outline of the region covered by our deep Parkes observations. Note that foreground emission from the Magellanic Stream as well as radio frequency interference were manually removed from a few channels of the data cube prior to creating the image.

\begin{figure}
  \includegraphics[width=\linewidth]{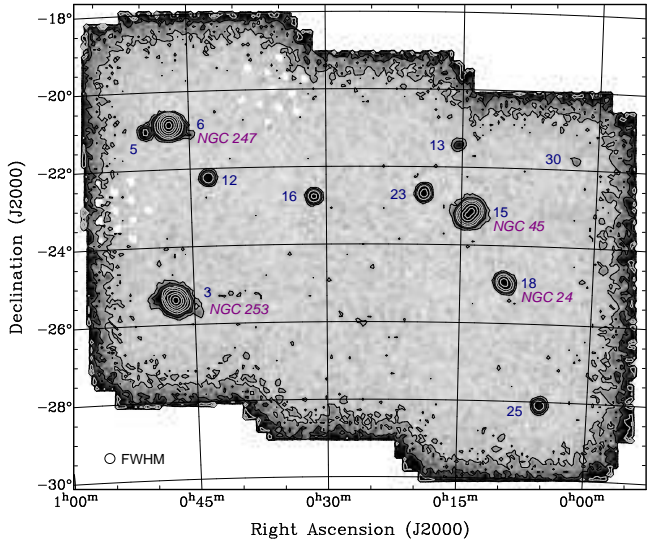}
  \caption{Peak flux density map of the northern Sculptor group region based on the combined HIPASS~2 and Parkes data (see outline in Fig.~\ref{fig_smax}). Contours are drawn at levels of $0.01$, $0.02$, $0.05$, $0.1$, $0.2$, $0.5$, $1$ and $2~\mathrm{Jy}$ per beam. The numbers correspond to the source ID in Tables~\ref{table_hipass_source_list} and~\ref{tab_parameters}. The four brightest galaxies have been labelled with their NGC number. The signal along the edges of the field is due to increasing noise. Some residual interference in the upper-left quadrant of the map has been flagged. Note that some of the extended features near the brightest galaxies are artefacts due to instrumental effects.}
  \label{fig_smax2}
\end{figure}

Lastly, we combined the data from our Parkes observations with HIPASS~2 to obtain an even deeper \ion{H}{I} data cube of the northern part of the Sculptor group. For this purpose, we first spectrally binned the Parkes data to $23.1~\mathrm{km \, s^{-1}}$ (approximately the velocity resolution of the HIPASS~2 data) and then regridded the HIPASS~2 data to the spatial and spectral grid of the Parkes data. The resulting data cube has an RMS noise level of just $2.5~\mathrm{mJy}$, which is an additional 8~per cent lower than the RMS of the deep Parkes data alone. The resolution and sensitivity of all three data products are summarised in Table~\ref{tab_data}. A peak flux density map created from the combined data cube is displayed in Fig.~\ref{fig_smax2}.

\begin{figure}
  \centering
  \includegraphics[width=0.87\linewidth]{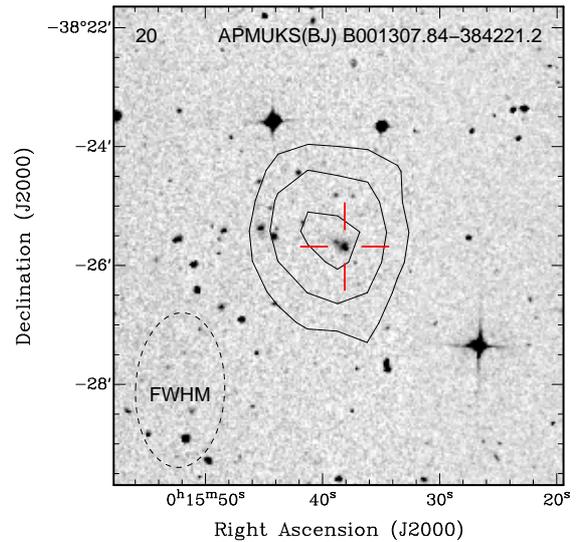}
  \caption{Optical image from the Digitized Sky Survey of the likely optical counterpart of source~20, APMUKS(BJ) B001307.84$-$384221.2 (marked with a red cross-hair), with \ion{H}{I} column density contours from ATCA data at levels of $2$, $4$ and $6 \times 10^{19}~\mathrm{cm}^{-2}$. The beam size of $154 \times 93~\mathrm{arcsec}$ of the ATCA data is indicated by the dashed ellipse. Note that the HIPASS~2 beam size of $15.5~\mathrm{arcmin}$ by far exceeds the size of the map shown here.}
  \label{fig_20_atca}
\end{figure}

\section{Data analysis}
\label{sect_analysis}

\subsection{Source extraction}

The HIPASS~2 data cube was searched for \ion{H}{I}~emission using the \textsc{Source Finding Application} (\textsc{SoFiA}; \citealt{Serra2015}).\footnote{\textsc{SoFiA} website: \texttt{https://github.com/SoFiA-Admin/SoFiA/}} We applied the smooth+clip algorithm with a range of four spectral and three spatial filter sizes and a detection threshold of $3 \, \sigma$. In addition, we made use of \textsc{SoFiA}'s reliability calculation based on negative detections \citep{Serra2012} and set a reliability threshold of 90~per cent. In total, \textsc{SoFiA} detected 80~sources across our survey volume, all of which were independently checked by eye. Of these 80~detections, 31~\ion{H}{I} sources were deemed to be reliable after visual inspection. We removed detections that were likely to be either artefacts (e.g. radio frequency interference) or -- based on their velocity and morphology -- foreground emission associated with the Milky Way or the Magellanic Stream.\footnote{Both interference and foreground emission would be classified as reliable by SoFiA, as they constitute genuine signals that are inconsistent with Gaussian noise.} The positions of all 31~reliable detections are marked and labelled in Fig.~\ref{fig_smax}.

\begin{table*}
  \caption{Comparison of integrated fluxes of several bright galaxies in our sample with previous measurements from the literature. The uncertainty is not listed in cases where no uncertainty was published. Note that interferometric observations with the VLA and ATCA may be affected by missing flux on large angular scales due to a lack of short baselines.}
  \label{tab_fluxcomparison}
  \begin{tabular}{lrll}
    \hline
    Name     & $F_{\rm int}$      & Telescope          & Reference \\
             & (Jy\,km\,s$^{-1}$) &                    &           \\
    \hline
    NGC~24   & $51 \pm \ph{00}5$  & Parkes 64~m        & \textit{this work} \\
             & $54 \pm \ph{00}5$  & VLA                & \citet{Chemin2006} \\
             & $50 \pm \ph{00}5$  & Parkes 64~m        & \citet{Koribalski2004} \\
             & $46 \pm \ph{00}4$  & Jodrell Bank 76~m  & \citet{Staveley-Smith1988} \\
             & $46 \pm \ph{00}3$  & Effelsberg 100~m   & \citet{Huchtmeier1985} \\
             & $74 \pm \ph{00}6$  & Parkes 64~m        & \citet{Reif1982} \\
    \hline
    NGC~45   & $206 \pm \ph{0}21$ & Parkes 64~m        & \textit{this work} \\
             & $186 \pm \ph{0}19$ & VLA                & \citet{Chemin2006} \\
             & $196 \pm \ph{0}14$ & Parkes 64~m        & \citet{Koribalski2004} \\
             & $242 \pm \ph{00}4$ & Effelsberg 100~m   & \citet{Huchtmeier1985} \\
             & $222 \pm \ph{0}10$ & Parkes 64~m        & \citet{Reif1982} \\
    \hline
    NGC~55   & $2105 \pm 211$     & Parkes 64~m        & \textit{this work} \\
             & $1980 \pm 100$     & ATCA               & \citet{Westmeier2013} \\
             & $1990 \pm 145$     & Parkes 64~m        & \citet{Koribalski2004} \\
             & $2450 \pm 250$     & Harvard 60~ft      & \citet{Eppstein1964} \\
    \hline
    NGC~247  & $607 \pm \ph{0}61$ & Parkes 64~m        & \textit{this work} \\
             & $608 \pm \ph{0}42$ & Parkes 64~m        & \citet{Koribalski2004} \\
             & $528 \pm \ph{0}18$ & VLA                & \citet{Carignan1990b} \\
             & $676 \pm \ph{00}6$ & Effelsberg 100~m   & \citet{Huchtmeier1985} \\
             & $530 \pm 160$      & Harvard 60~ft      & \citet{Eppstein1964} \\
    \hline
    NGC~253  & $668 \pm \ph{0}67$ & Parkes 64~m        & \textit{this work} \\
             & $693 \pm \ph{0}40$ & Parkes 64~m        & \citet{Koribalski2004} \\
             & $756 \pm \ph{0}74$ & Jodrell Bank Mk~II & \citet{Staveley-Smith1988} \\
             & $795 \pm \ph{00}9$ & Effelsberg 100~m   & \citet{Huchtmeier1985} \\
    \hline
    NGC~300  & $2083 \pm 208$     & Parkes 64~m        & \textit{this work} \\
             & $1720 \: \ph{\pm} \: \ph{000}$ & ATCA   & \citet{Westmeier2011} \\
             & $1973 \pm 156$     & Parkes 64~m        & \citet{Koribalski2004} \\
             & $2010 \pm 200$     & Harvard 60~ft      & \citet{Eppstein1964} \\
    \hline
    NGC~625  & $33 \pm \ph{00}3$  & Parkes 64~m        & \textit{this work} \\
             & $31 \pm \ph{00}4$  & Parkes 64~m        & \citet{Koribalski2004} \\
             & $31 \: \ph{\pm} \: \ph{000}$ & Parkes 64~m & \citet{Meyer2004} \\
             & $38 \pm \ph{00}4$  & Parkes 64~m        & \citet{Reif1982} \\
    \hline
    NGC~7793 & $287 \pm \ph{0}29$ & Parkes 64~m        & \textit{this work} \\
             & $279 \pm \ph{0}20$ & Parkes 64~m        & \citet{Koribalski2004} \\
             & $281 \: \ph{\pm} \: \ph{000}$ & Parkes 64~m & \citet{Meyer2004} \\
             & $268 \pm \ph{0}15$ & Parkes 64~m        & \citet{Reif1982} \\
    \hline
  \end{tabular}
\end{table*}

Next, we cross-matched our detections with those from the HIPASS catalogue \citep{Meyer2004}, the HIPASS Bright Galaxy Catalog \citep{Koribalski2004} as well as the dwarf galaxy \ion{H}{I}~detections in the Sculptor group published by \citet{Cote1997} and \citet{Bouchard2005}, all of which are marked in Fig.~\ref{fig_smax} for comparison. Eight out of our 31~detections (more than a quarter, flagged with `n' in Table~\ref{table_hipass_source_list}) are new and -- to the best of our knowledge -- have never before been detected in \ion{H}{I}~emission. All other sources are included in the main HIPASS catalogue or the HIPASS Bright Galaxy Catalog, with the notable exception of NGC~59 \citep{Beaulieu2006}. Three sources from the HIPASS catalogue and two detections reported by \citet{Cote1997} are not marked as detections in Fig.~\ref{fig_smax}. While these five sources, located near $\alpha = 0^{\mathrm{h}}$ and $\delta = -40\degr{}$, are clearly visible in our data, they are almost certainly extended gas clouds associated with the Magellanic Stream rather than galaxies in the Sculptor filament; hence they are not included in our sample. Curiously, none of the four dwarf galaxies found in \ion{H}{I}~emission by \citet{Bouchard2005} is listed in our catalogue, although our sensitivity should have been sufficient to clearly detect at least the brightest objects. A closer investigation reveals that in the case of ESO~410$-$G005 we do see a faint signal in our HIPASS~2 data, but the emission appears to be part of a more extended component of the Magellanic Stream rather than spatially unresolved emission from the galaxy itself. In the case of ESO~540$-$G030 there is a tentative signal at the correct position and velocity, but the emission is too faint to be considered statistically significant and therefore not included in our catalogue. At the positions of the two other galaxies, ESO~294$-$G010 and ESO~540$-$G032, there is no hint of any emission in our data, possibly due to insufficient sensitivity.

\begin{table*}
  \caption{List of \ion{H}{I}~sources found in the HIPASS~2 data cube of the Sculptor group region. The columns denote the identification number (ID), flag indicating that the \ion{H}{I} detection is new (n) and no other redshift measurement is available (r), name of the most likely optical counterpart, HIPASS cross-identification, right ascension ($\alpha$) and declination ($\delta$), barycentric velocity ($\mathrm{c}z$), line width ($w_{50}$), integrated flux ($F_{\rm int}$), distance ($d$; see footnotes for origin), and \ion{H}{I} mass ($M_{\ion{H}{I}}$). The uncertainties in $\mathrm{c}z$ are of the order of $5~\mathrm{km \, s}^{-1}$, while the uncertainties in $w_{50}$ and $F_{\rm int}$ are about $8~\mathrm{km \, s}^{-1}$ and 10~per cent, respectively. Note that the tabulated line widths have not yet been corrected for the finite spectral resolution of HIPASS~2 of $\Delta v \approx 26.4~\mathrm{km \, s}^{-1}$.}
  \label{table_hipass_source_list}
  \begin{tabular}{rlp{1.8cm}lrrrrrrr}
     \hline
     ID & flag & optical        & HIPASS      & $\alpha$ & $\delta$  & $\mathrm{c}z$             &  $w_{50}$                 & $\log_{10}(F_{\rm int}$         & $d$   & $\log_{10}(M_{\ion{H}{I}}$ \\
        &      & counterpart    &             & (J2000)  & (J2000)   & ($\mathrm{km \, s}^{-1}$) & ($\mathrm{km \, s}^{-1}$) & $/\mathrm{Jy \, km \, s}^{-1})$ & (Mpc) & $/\mathrm{M}_{\sun})$ \\
     \hline
      1 &      & NGC 55         & J0015$-$39  & 00:15:11 & $-$39:13:16 &  134.0 & 165.8 & $ 3.323$ & $ 2.0 \pm 0.3^{\rm a}$       & $9.28_{-0.19}^{+0.17}$ \\
      2 &      & NGC 300        & J0054$-$37  & 00:54:53 & $-$37:41:04 &  146.6 & 149.8 & $ 3.319$ & $ 2.0 \pm 0.3^{\rm a}$       & $9.28_{-0.17}^{+0.15}$ \\
      3 &      & NGC 253        & J0047$-$25  & 00:47:29 & $-$25:17:34 &  255.2 & 415.3 & $ 2.825$ & $ 3.2 \pm 0.5^{\rm a}$       & $9.20_{-0.21}^{+0.18}$ \\
      4 &      & PGC 621        & J0008$-$34  & 00:08:13 & $-$34:34:25 &  220.3 &  33.1 & $ 0.651$ & $ 3.2 \pm 0.5^{\rm a}$       & $7.04_{-0.20}^{+0.18}$ \\
      5 &      & UGCA 15        & J0049$-$20  & 00:49:45 & $-$21:00:18 &  295.4 &  33.1 & $ 0.548$ & $ 3.4 \pm 0.2^{\rm a}$       & $6.99_{-0.09}^{+0.09}$ \\
      6 &      & NGC 247        & J0047$-$20  & 00:47:06 & $-$20:44:45 &  151.2 & 201.7 & $ 2.783$ & $ 3.6 \pm 0.5^{\rm a}$       & $9.27_{-0.17}^{+0.15}$ \\
      7 & nr   & \sourceSV      & --          & 00:55:07 & $-$23:12:22 &  249.6 &  54.6 & $ 0.154$ & $ 3.6 \pm 2.1^{\rm e}$       & $6.63_{-0.82}^{+0.44}$ \\
      8 &      & NGC 625        & J0135$-$41  & 01:35:05 & $-$41:26:17 &  394.9 &  72.8 & $ 1.518$ & $ 3.8 \pm 0.6^{\rm a}$       & $8.05_{-0.18}^{+0.16}$ \\
      9 &      & NGC 7793       & J2357$-$32  & 23:57:29 & $-$32:34:29 &  223.8 & 175.6 & $ 2.458$ & $ 4.1 \pm 0.7^{\rm a}$       & $9.05_{-0.22}^{+0.18}$ \\
     10 &      & PGC 6430       & J0145$-$43  & 01:45:01 & $-$43:36:16 &  391.7 &  62.4 & $ 1.903$ & $ 4.2 \pm 0.5^{\rm a}$       & $8.52_{-0.16}^{+0.14}$ \\
     11 &      & UGCA 442       & J2343$-$31  & 23:43:40 & $-$31:58:11 &  266.7 &  93.8 & $ 1.749$ & $ 4.3_{-0.5}^{+0.6 {\rm b}}$ & $8.38_{-0.15}^{+0.15}$ \\
     12 &      & IC 1574        & J0043$-$22  & 00:43:01 & $-$22:12:36 &  363.3 &  40.7 & $ 0.688$ & $ 4.8 \pm 0.2^{\rm a}$       & $7.42_{-0.09}^{+0.08}$ \\
     13 &      & NGC 59         & --          & 00:15:09 & $-$21:24:00 &  365.1 &  44.5 & $ 0.207$ & $ 4.8 \pm 0.6^{\rm a}$       & $6.95_{-0.17}^{+0.15}$ \\
     14 & n    & \sourceXCIX    & --          & 23:42:04 & $-$33:08:43 &  459.0 &  48.3 & $-0.102$ & $ 6.6 \pm 2.1^{\rm e}$       & $6.90_{-0.38}^{+0.28}$ \\
     15 &      & NGC 45         & J0014$-$23  & 00:14:04 & $-$23:11:01 &  468.0 & 168.5 & $ 2.313$ & $ 6.6 \pm 0.2^{\rm c}$       & $9.33_{-0.08}^{+0.07}$ \\
     16 &      & PGC 1920       & J0031$-$22  & 00:31:21 & $-$22:46:04 &  539.0 &  41.9 & $ 0.803$ & $ 7.6 \pm 2.1^{\rm e}$       & $7.95_{-0.32}^{+0.25}$ \\
     17 &      & PGC 71464      & J2326$-$37  & 23:26:54 & $-$37:21:20 &  693.2 &  69.6 & $ 0.944$ & $ 7.9 \pm 0.4^{\rm a}$       & $8.11_{-0.09}^{+0.09}$ \\
     18 &      & NGC 24         & J0009$-$24  & 00:09:55 & $-$24:57:59 &  551.2 & 212.7 & $ 1.710$ & $ 8.2 \pm 1.3^{\rm a}$       & $8.91_{-0.19}^{+0.17}$ \\
     19 &      & IC 5332        & J2334$-$36  & 23:34:29 & $-$36:05:40 &  701.5 & 102.7 & $ 2.217$ & $ 8.4_{-1.4}^{+1.7 {\rm d}}$ & $9.44_{-0.20}^{+0.20}$ \\
     20 & nr   & \sourceXLIV    & --          & 00:15:10 & $-$38:21:32 &  606.0 &  58.6 & $ 0.128$ & $ 8.7 \pm 2.1^{\rm e}$       & $7.38_{-0.29}^{+0.23}$ \\
     21 &      & PGC 72525      & J2349$-$37  & 23:49:31 & $-$37:46:34 &  646.5 &  86.4 & $ 1.087$ & $ 9.2 \pm 2.1^{\rm e}$       & $8.39_{-0.27}^{+0.22}$ \\
     22 & n    & LEDA 166061    & --          & 01:08:22 & $-$38:13:09 &  654.5 &  32.0 & $ 0.147$ & $ 9.3 \pm 2.1^{\rm e}$       & $7.46_{-0.27}^{+0.22}$ \\
     23 &      & PGC 1242       & J0019$-$22  & 00:19:13 & $-$22:39:23 &  669.2 & 117.8 & $ 1.203$ & $ 9.6 \pm 2.1^{\rm e}$       & $8.54_{-0.26}^{+0.21}$ \\
     24 &      & NGC 7713       & J2336$-$37a & 23:36:15 & $-$37:56:37 &  695.5 & 186.3 & $ 1.757$ & $10.0 \pm 1.5^{\rm a}$       & $9.13_{-0.18}^{+0.16}$ \\
     25 &      & PGC 388        & J0005$-$28  & 00:05:37 & $-$28:06:16 &  740.7 &  47.7 & $ 0.701$ & $10.6 \pm 2.1^{\rm e}$       & $8.12_{-0.24}^{+0.20}$ \\
     26 & nr   & LEDA 166060    & --          & 00:37:42 & $-$43:27:19 &  783.3 &  29.2 & $-0.102$ & $11.2 \pm 2.1^{\rm e}$       & $7.37_{-0.23}^{+0.19}$ \\
     27 &      & NGC 244        & J0045$-$15  & 00:45:45 & $-$15:35:08 &  942.6 &  54.4 & $ 0.676$ & $11.6_{-3.5}^{+5.2 {\rm d}}$ & $8.18_{-0.36}^{+0.36}$ \\
     28 & nr   & \sourceCLXXXV  & --          & 00:14:11 & $-$17:48:53 &  819.9 &  51.0 & $-0.038$ & $11.7 \pm 2.1^{\rm e}$       & $7.47_{-0.22}^{+0.18}$ \\
     29 & n    & LEDA 198121    & --          & 23:58:42 & $-$22:46:08 &  952.6 &  85.7 & $ 0.079$ & $13.6 \pm 2.1^{\rm e}$       & $7.72_{-0.19}^{+0.17}$ \\
     30 & nr   & \sourceCLXXVII & --          & 00:02:22 & $-$21:48:00 &  987.6 & 118.3 & $ 0.351$ & $14.1 \pm 2.1^{\rm e}$       & $8.02_{-0.19}^{+0.16}$ \\
     31 &      & LEDA 812517    & J2349$-$22  & 23:49:43 & $-$22:33:59 & 1030.8 &  76.2 & $ 0.308$ & $14.7 \pm 2.1^{\rm e}$       & $8.02_{-0.18}^{+0.16}$ \\
    \hline
    \multicolumn{11}{l}{$^{\rm a} \, $Median $\pm$ standard deviation of distances listed in NASA/IPAC Extragalactic Database.} \\
    \multicolumn{11}{l}{$^{\rm b} \, $TRGB distance determined by \citet{Karachentsev2003}.} \\
    \multicolumn{11}{l}{$^{\rm c} \, $TRGB distance determined by \citet{Jacobs2009}.} \\
    \multicolumn{11}{l}{$^{\rm d} \, $Distance from Nearby Galaxies Catalog \citep{Tully1988}.} \\
    \multicolumn{11}{l}{$^{\rm e} \, $No distance measurement available; assuming $d = \mathrm{c} z / H_{0}$, where $H_{0} = 70~\mathrm{km \, s^{-1} \, Mpc^{-1}}$.} \\
  \end{tabular}
\end{table*}

\ion{H}{I}~contours of all detections on top of optical images from the Digital Sky Survey (DSS) are presented in Appendix~\ref{app_maps}. In all 31~cases we see a potential optical counterpart in the DSS data, suggesting that all 31~\ion{H}{I} detections are galaxies with a significant stellar population. While most of the detected galaxies are well-known and have optical redshift and distance measurements, there are five potential optical counterparts without any redshift information in the NASA/IPAC Extragalactic Database (flagged with `r' in Table~\ref{table_hipass_source_list}). Assuming that the detected \ion{H}{I}~emission is indeed associated with those optical counterparts, this would confirm for the first time that they are nearby dwarf galaxies in the Sculptor filament rather than distant background galaxies at higher redshift.

Due to a fortunate coincidence, one of the five objects without optical redshift information, source number~20 in our catalogue, happened to be covered by the ATCA observations of NGC~55 presented by \citet{Westmeier2013}. ATCA \ion{H}{I} column density contours of this source on top of an optical image from the Digitized Sky Survey are presented in Fig.~\ref{fig_20_atca}. At a much higher angular resolution of $154 \times 93~\mathrm{arcsec}$, the ATCA data confirm that the \ion{H}{I} emission indeed is spatially coincident with the optical source \sourceXLIV, and a physical association appears highly likely even without optical redshift confirmation. Additional evidence stems from the fact that the \ion{H}{I} emission appears marginally extended in the same direction as the major axis of the clumpy optical emission.

While optical and \ion{H}{I} redshifts agree within a few $\mathrm{km \, s}^{-1}$ for most of our detections, there are two cases, (14) 2dFGRS~S491Z111 and (29) LEDA~198121, where we find a discrepancy between the optical and \ion{H}{I} redshift measurements of about $50~\mathrm{km \, s}^{-1}$. In both cases, the discrepancy is caused by large uncertainties in the optical redshift, and both values are in excellent agreement within those uncertainties. Consequently, our new \ion{H}{I} redshifts for those two sources are far more accurate and precise than the existing optical redshifts.

\begin{table*}
  \caption{Observational and physical parameters of the eleven \ion{H}{I}~sources detected in our Parkes northern Sculptor group survey. The columns denote the identification number (ID), name of the potential optical counterpart, HIPASS cross-identification, right ascension ($\alpha$) and declination ($\delta$), barycentric velocity ($\mathrm{c}z$), line width ($w_{50}$), integrated flux ($F_{\rm int}$), distance ($d$; see footnotes for origin), and \ion{H}{I} mass ($M_{\ion{H}{I}}$). The uncertainties in $\mathrm{c}z$ are of the order of $5~\mathrm{km \, s}^{-1}$, while $w_{50}$ and $F_{\rm int}$ are accurate to within about $8~\mathrm{km \, s}^{-1}$ and 10~per cent, respectively. While the tabulated line widths have not yet been corrected for the finite spectral resolution of our Parkes data of $\Delta v \approx 1.6~\mathrm{km \, s}^{-1}$, the correction factor will be negligible, as all line profiles are well-resolved.}
  \label{tab_parameters}
  \begin{tabular}{rp{1.8cm}lrrrrrrr}
    \hline
    ID & optical        & HIPASS     & $\alpha$ & $\delta$    & c$z$           & $w_{50}$       & $\log_{10}(F_{\rm int}$         & $d$                    & $\log_{10}(M_{\ion{H}{I}}$ \\
       & counterpart    &            & (J2000)  & (J2000)     & (km\,s$^{-1}$) & (km\,s$^{-1}$) & $/\mathrm{Jy \, km \, s}^{-1})$ & (Mpc)                  & $/\mathrm{M}_{\sun})$ \\
    \hline
     3 & NGC 253        & J0047$-$25 & 00:47:10 & $-$25:21:39 & $260.6$        & $409.5$        & $2.839$                         &  $3.3 \pm 0.5^{\rm a}$ & $9.24_{-0.21}^{+0.18}$ \\
     5 & UGCA 15        & J0049$-$20 & 00:49:45 & $-$21:01:01 & $295.3$        & $ 19.3$        & $0.415$                         &  $3.4 \pm 0.2^{\rm a}$ & $6.85_{-0.09}^{+0.09}$ \\
     6 & NGC 247        & J0047$-$20 & 00:47:10 & $-$20:49:24 & $153.0$        & $199.0$        & $2.781$                         &  $3.7 \pm 0.5^{\rm a}$ & $9.28_{-0.17}^{+0.15}$ \\
    12 & IC 1574        & J0043$-$22 & 00:43:01 & $-$22:14:01 & $362.1$        & $ 41.1$        & $0.732$                         &  $4.9 \pm 0.2^{\rm a}$ & $7.48_{-0.09}^{+0.08}$ \\
    13 & NGC 59         & --         & 00:15:22 & $-$21:25:19 & $367.6$        & $ 45.4$        & $0.362$                         &  $4.8 \pm 0.6^{\rm a}$ & $7.11_{-0.17}^{+0.15}$ \\
    15 & NGC 45         & J0014$-$23 & 00:14:06 & $-$23:11:24 & $468.3$        & $166.5$        & $2.344$                         &  $6.6 \pm 0.2^{\rm b}$ & $9.36_{-0.08}^{+0.07}$ \\
    16 & PGC 1920       & J0031$-$22 & 00:31:22 & $-$22:46:21 & $541.7$        & $ 33.3$        & $0.756$                         &  $7.7 \pm 2.1^{\rm c}$ & $7.91_{-0.32}^{+0.25}$ \\
    18 & NGC 24         & J0009$-$24 & 00:09:58 & $-$24:57:21 & $553.7$        & $206.7$        & $1.653$                         &  $7.8 \pm 1.2^{\rm a}$ & $8.81_{-0.19}^{+0.17}$ \\
    23 & PGC 1242       & J0019$-$22 & 00:19:11 & $-$22:39:57 & $670.0$        & $113.4$        & $1.104$                         &  $9.6 \pm 2.1^{\rm c}$ & $8.44_{-0.26}^{+0.21}$ \\
    25 & PGC 388        & J0005$-$28 & 00:05:32 & $-$28:05:35 & $737.6$        & $ 35.0$        & $0.653$                         & $10.6 \pm 2.1^{\rm c}$ & $8.07_{-0.24}^{+0.20}$ \\
    30 & \sourceCLXXVII & --         & 00:02:28 & $-$21:47:49 & $969.5$        & $ 88.4$        & $0.279$                         & $13.9 \pm 2.1^{\rm c}$ & $7.94_{-0.19}^{+0.16}$ \\
    \hline
    \multicolumn{10}{l}{$^{\rm a} \, $Median $\pm$ standard deviation of distances listed in NASA/IPAC Extragalactic Database.} \\
    \multicolumn{10}{l}{$^{\rm b} \, $TRGB distance determined by \citet{Jacobs2009}.} \\
    \multicolumn{10}{l}{$^{\rm c} \, $No distance measurement available; assuming $d = \mathrm{c} z / H_{0}$, where $H_{0} = 70~\mathrm{km \, s^{-1} \, Mpc^{-1}}$.} \\
  \end{tabular}
\end{table*}

\subsection{Source characterisation}

As all of our galaxies are only marginally resolved by the gridded $15.5$-arcmin Parkes beam, we generated integrated spectra of all detected sources for the purpose of parameterisation, using the \textsc{Miriad} \citep{Sault1995} task \textsc{mbspect}. In each case, a low-order polynomial was fitted to the spectral baseline in order to remove any residual baseline variations from the spectrum prior to parameterisation. We then fitted the so-called Busy Function \citep{Westmeier2014} to the spectral profile of each galaxy for the purpose of parameterisation. The integrated spectra and Busy Function fits of all detected galaxies are presented in Appendix~\ref{app_spectra}. From the free parameters of the fit we then derived relevant observational parameters for each galaxy, including the barycentric velocity centroid, $\mathrm{c}z$, the line width at 50~per cent of the peak flux density, $w_{50}$, and the integrated flux, $F_{\rm int}$. The results of our parameterisation are presented in Table~\ref{table_hipass_source_list} (for our HIPASS~2 data) and Table~\ref{tab_parameters} (for our Parkes survey of the northern Sculptor group).

In principle, the Busy Function fit would also provide us with statistical uncertainties for all measured parameters. However, these statistical uncertainties significantly underestimate the true uncertainties of our measurements, which are dominated by systematic effects rather than statistical noise. Such effects include flux calibration errors, deviation of the beam from the assumed $15.5$-arcmin Gaussian, errors resulting from baseline fitting and subtraction, errors introduced in the process of creating the integrated spectrum, and the effects of radio-frequency interference. Simple tests demonstrated that these systematic effects exceed the statistical uncertainties of our measurements by more than one order of magnitude. Consequently, we refrain from specifying statistical uncertainties in Table~\ref{tab_parameters}, as their small values would be misleading. Instead, we estimate that the systematic uncertainties in the frequency centroid measurements are of the order of $25~\mathrm{kHz}$ (equivalent to about $5~\mathrm{km \, s}^{-1}$), while line width and integrated flux measurements are accurate to within about $8~\mathrm{km \, s}^{-1}$ and 10~per cent, respectively.

Redshift-independent distances, where available, were extracted from the NASA/IPAC Extragalactic Database (NED). We usually adopted the median value of the distances listed in NED, with the uncertainty assumed to be the standard deviation. Note that we excluded obvious outliers before calculating the median, and our tabulated distances may therefore be different from the ones automatically generated by NED. In a few cases the distance measurements listed in NED were inconsistent with each other, and we instead adopted the most reliable distance measurement (usually the TRGB distance, where available). Finally, for many galaxies redshift-independent distance measurements were unavailable. In these cases we assumed a distance of $d = \mathrm{c}z / H_{0}$, with $H_{0} = 70~\mathrm{km \, s^{-1} \, Mpc^{-1}}$ \citep{Bennett2013}, and an uncertainty of $\sigma_{v} / H_{0} = 2.1~\mathrm{Mpc}$ under the assumption that $\sigma_{v} = 150~\mathrm{km \, s}^{-1}$, as suggested by the mean velocity dispersion of nearby galaxy groups in the catalogue of \citet{Huchra1982}. Note that we refrain from converting our barycentric velocities into the Local Group standard-of-rest for this purpose, as both the GSR and LGSR velocity corrections in the direction of the Sculptor group are significantly smaller (typically $\lesssim 50~\mathrm{km \, s}^{-1}$) than the statistical uncertainty introduced by the group velocity dispersion ($\approx 150~\mathrm{km \, s}^{-1}$). Hence, there is no benefit in carrying out higher-order rest-frame corrections which would only introduce additional systematic errors, as neither the GSR nor the LGSR are constrained to better than about $30~\mathrm{km \, s}^{-1}$ \citep{Einasto1982,vanderMarel2012,Schoenrich2012}.

\begin{figure*}
	\includegraphics[width=\linewidth]{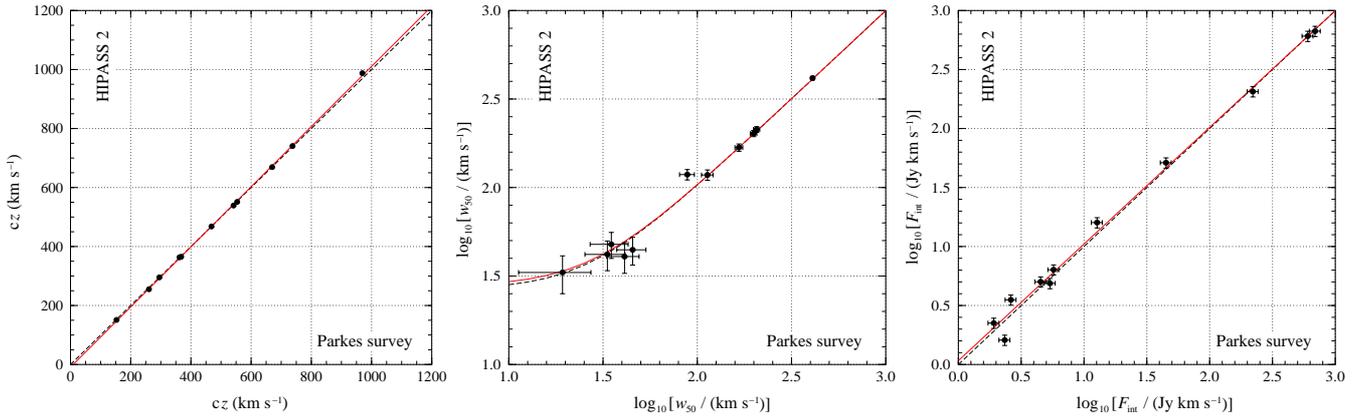}
	\caption{Comparison of recession velocity (left), $w_{50}$ line width (centre) and integrated flux (right) between our Parkes survey (abscissa) and HIPASS~2 (ordinate) for the eleven sources that are present in both data sets. The dashed, black line represents our theoretical expectation, while the red line depicts the result of a fit to the data. We fitted $f(x) = a x + b$ in the case of $\mathrm{c}z$, $f(x) = \sqrt{x^{2} + a^{2}}$ in the case of $w_{50}$, and $f(x) = a \log_{10}(x) + b$ in the case of $F_{\rm int}$, where $a$ and $b$ are free parameters of the fit.}
	\label{fig_comparison}
\end{figure*}

\subsection{Comparison of observational parameters}

Eleven \ion{H}{I}~sources have been detected in both our deep Parkes survey of the northern Sculptor group and in the HIPASS~2 data of the entire Sculptor group. In Fig.~\ref{fig_comparison} we compare the measurements of radial velocity, $w_{50}$ line width and integrated flux between the two data sets. In all three cases we find a generally good agreement between the measured values and our expectation, indicating that both the flux calibration and frequency calibration of the data sets are consistent. Note that in the case of $w_{50}$ we do not expect to see a linear relation due to the substantially worse velocity resolution of HIPASS~2 ($\Delta v = 26.4~\mathrm{km \, s}^{-1}$ after Hann-filtering) compared to $1.6~\mathrm{km \, s}^{-1}$ for our Parkes data. Hence, all line width measurements in HIPASS~2 will need to be deconvolved to recover the true line width.

In order to ensure that our flux calibration is correct, we also compare the fluxes measured in our HIPASS~2 data with fluxes from the literature for several bright galaxies in our sample (Table~\ref{tab_fluxcomparison}). In general, we find an excellent agreement between integrated flux measurements, in particular those made with the Parkes telescope, suggesting that the flux calibration of both our HIPASS~2 and deep Parkes observations is accurate. In a few cases, flux measurements appear to be inconsistent with one another, in particular when different telescopes were involved. The cause of these discrepancies, e.g.\ between measurements made with Effelsberg and Parkes, is not entirely clear. Possible explanations include insufficient sky coverage of older observations due to a lack of sensitivity, errors related to standing waves or baseline fitting, and systematic differences in flux calibration between observatories in the northern and southern hemisphere due to the use of different calibrator sources (see, e.g., \citealt{Kalberla2015}). In addition, the uncertainties published by \citet{Huchtmeier1985} for their Effelsberg detections are not credible and in some cases smaller than the expected flux calibration accuracy of the telescope. Lastly, interferometric flux measurements are expected to be lower due to the possibility of missing flux on large angular scales.

\section{Results}
\label{sect_results}

\subsection{Properties of \ion{H}{I} detections}

The observational and physical parameters of the 31~sources detected in our survey are listed in Tables~\ref{table_hipass_source_list} (HIPASS~2) and~\ref{tab_parameters} (deep Parkes survey of the northern Sculptor group). Our detections cover a barycentric redshift range of $\mathrm{c}z \approx 100$ to $1000~\mathrm{km \, s}^{-1}$. The far end corresponds to a Hubble distance of $\mathrm{c}z / H_{0} \approx 14.3~\mathrm{Mpc}$, beyond which the Sculptor filament appears to fade away. Hence, throughout this paper we will only consider objects in the distance range of $1.7 \le d \le 15~\mathrm{Mpc}$, which encompasses the entire Sculptor filament from the nearby NGC~55 group to the far end of the filament.

The $w_{50}$ line widths of our detections range from about $20$ to $200~\mathrm{km \, s}^{-1}$. The only exception is NGC~253 which has a much larger line width of just over $400~\mathrm{km \, s}^{-1}$. Note that the tabulated line widths are the raw widths measured from the integrated spectrum and have not yet been corrected for the effects of spectral resolution and disc inclination.

The \ion{H}{I} masses of our detections cover three orders of magnitude from $4 \times 10^{6}$ to $3 \times 10^{9}~\mathrm{M}_{\sun}$, with a $10 \, \sigma$ mass detection threshold of about $10^{6}~\mathrm{M}_{\sun}$ at the near end ($\approx 2~\mathrm{Mpc}$) and $6 \times 10^{7}~\mathrm{M}_{\sun}$ at the far end ($\approx 15~\mathrm{Mpc}$) of the Sculptor filament. Hence, even the most massive galaxies in our sample have \ion{H}{I} masses below $M_{\ion{H}{I}}^{\star}$ \citep{Zwaan2005,Martin2010}, and our sample does not extend to the turnover point of the HIMF.

\begin{figure}
	\includegraphics[width=\linewidth]{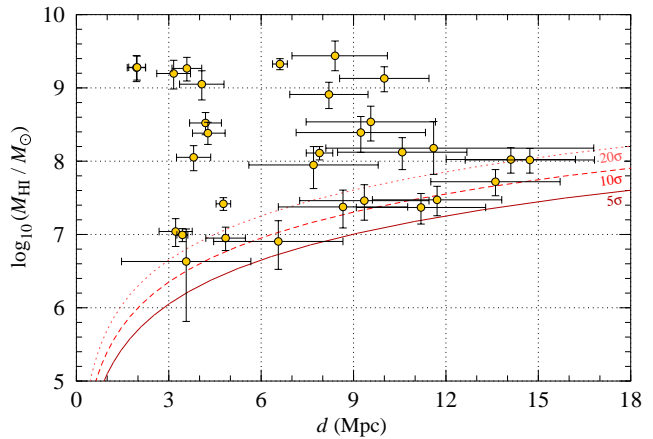}
	\caption{\ion{H}{I}~mass versus distance of the galaxies detected in our survey. The solid, dashed, and dotted red curves show our $5$, $10$, and $20 \, \sigma$~detection limits, respectively, under the assumption of a constant line width of $26.4~\mathrm{km \, s}^{-1}$.}
	\label{fig_m-d}
\end{figure}

A plot of \ion{H}{I} mass versus distance of all detections is presented in Fig.~\ref{fig_m-d}, which also includes $5$, $10$ and $20 \, \sigma$ sensitivity curves over $20~\mathrm{km \, s}^{-1}$ for comparison. To our surprise, the density of galaxies does not appear to decrease with increasing \ion{H}{I} mass, but rather remains relatively constant across three orders of magnitude in \ion{H}{I} mass. In particular, more than a quarter of all our detections have \ion{H}{I} masses in excess of $10^{9}~\mathrm{M}_{\sun}$. This almost flat distribution of \ion{H}{I} masses is unexpected and likely in tension with measurements of the global HIMF \citep{Zwaan2005,Martin2010} from which we would expect a decrease in the number density of galaxies towards higher \ion{H}{I} mass. It is also remarkable that we do not detect any additional galaxies in our deep Parkes survey of the northern Sculptor group region even though the noise level is significantly lower than in the HIPASS~2 data. This unexpected result suggests that the mass distribution essentially remains flat below $10^{7}~\mathrm{M}_{\sun}$.

\subsection{\ion{H}{I} mass function}
\label{sect_himf}

The HIMF is usually expressed as the space density of galaxies per logarithmic mass interval and fitted with a Schechter function \citep{Schechter1976} of the form
\begin{equation}
	\Theta(\mu) \, \mathrm{d}\mu = \Theta^{\star} \mu^{\alpha} \exp(-\mu) \, \mathrm{d}\mu \\
\end{equation}
or, in logarithmic representation,
\begin{equation}
	\Phi(\mu) \, \mathrm{d} \! \log_{10}(\mu) = \ln(10) \, \Phi^{\star} \mu^{\alpha + 1} \exp(-\mu) \, \mathrm{d} \! \log_{10}(\mu)
\end{equation}
with the dimensionless parameter $\mu \equiv M_{\ion{H}{I}} / M_{\ion{H}{I}}^{\star}$, where $M_{\ion{H}{I}}$ is the \ion{H}{I}~mass, $M_{\ion{H}{I}}^{\star}$ is the characteristic mass corresponding to the turnover point of the mass function, $\Theta^{\star}$ or $\Phi^{\star}$ is the global normalisation factor, and $\alpha$ is the exponent of the power law that describes the slope, $\alpha + 1$, of the low-mass end of the HIMF in the commonly used logarithmic representation.

A comprehensive discussion of different methods of measuring the HIMF as well as their advantages and drawbacks can be found in \citet{Zwaan2003}. The simplest method of determining the HIMF is the $\sum V_{\rm max}^{-1}$ method \citep{Schmidt1968}. It is based on summing in each mass bin the inverse of the maximum volume across which the galaxies in that bin could still be detected within the boundaries and sensitivity of the survey. A more advanced method is the two-dimensional stepwise maximum likelihood (2DSWML) method developed by \citet{Zwaan2003}, which is less susceptible to the influence of cosmic variance than the $\sum V_{\rm max}^{-1}$ method. However, as it requires two-dimensional binning in \ion{H}{I}~mass and spectral line width, the 2DSWML method can not be applied in our case due to the small number of detections in our survey. We therefore use the $\sum V_{\rm max}^{-1}$ method here, given that cosmic variance is not of concern when studying an individual structure like the Sculptor filament.

To calculate the maximum detection volume for a particular galaxy in our sample, we must first define the maximum distance out to which a galaxy of \ion{H}{I}~mass $M_{\ion{H}{I}}$ could still be detected. Assuming an RMS noise level of $\sigma_{0}$ at the original velocity resolution of $\Delta v_{0}$, we can estimate the distance, $d_{\rm max}$, at which a spatially unresolved $M_{\ion{H}{I}}$ galaxy with a profile width of $\Delta v$ can be detected at the $6 \, \sigma$~level (integrated) to be
\begin{equation}
  d_{\rm max}^{2} = \frac{M_{\ion{H}{I}}}{6 \sigma_{0} \mathrm{C} \sqrt{\Delta v_{0} \Delta v}}
\end{equation}
where $\mathrm{C} \approx 2.36 \times 10^{5}~\mathrm{\mathrm{M}_{\sun} \, Mpc^{-2} \, Jy^{-1} \, km^{-1} \, s}$ is a constant. Our choice of a $6 \, \sigma$ threshold is slightly more conservative than the $5 \, \sigma$ threshold assumed in previous studies (e.g., \citealt{Zwaan1997}) and motivated by the fact that our faintest detections have integrated signal-to-noise ratios in the range of about $5.5$ to $6.5$. Note that for massive galaxies of $M_{\ion{H}{I}} \gtrsim 10^{8}~\mathrm{M}_{\sun}$ the value of $d_{\rm max}$ is usually limited by the bandwidth covered by the survey, as such galaxies would be detectable well beyond the boundary of the survey volume. In this case we set $d_{\rm max} = 15~\mathrm{Mpc}$ in accordance with the approximate distance of the far end of the Sculptor filament.

\begin{figure*}
	\includegraphics[width=0.7\linewidth]{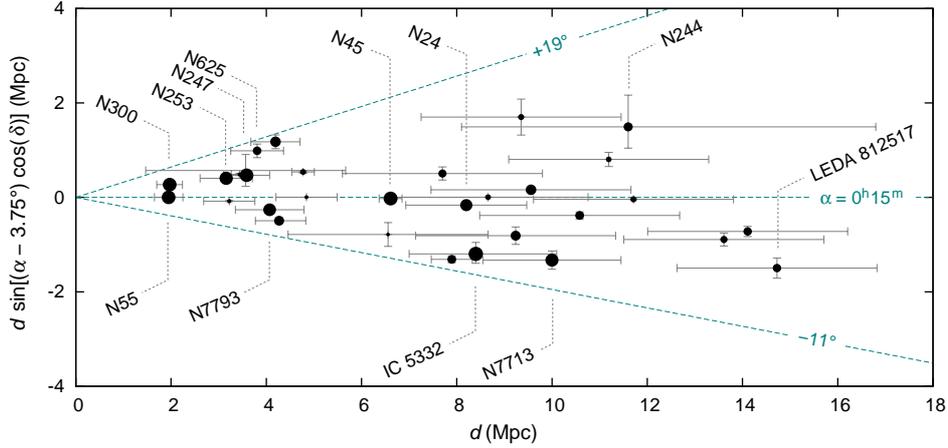}
	\caption{Plot of physical height above a plane of constant right ascension of $\alpha = 0^{\rm h}15^{\rm m}$ versus distance for the 31~\ion{H}{I} sources detected in our survey. The size of the data points is indicative of \ion{H}{I} mass. The dashed, teal lines indicate directions of constant values of $\Delta \alpha / \cos(\delta)$, with the outer envelopes marking the approximate boundaries of our survey area spanning an angular size of about $30\degr$. Several prominent galaxies are labelled with their NGC number or alternative designation.}
	\label{fig_filament}
\end{figure*}

Once the maximum distance is known, we then need to calculate the corresponding maximum volume. This step is far more complex, as it depends not only on the geometry of our survey area, but also on the actual spatial distribution of galaxies within the Sculptor filament. As discussed earlier and illustrated in Fig.~\ref{fig_filament}, the Sculptor filament is roughly oriented along the line of sight with an apparent thickness of approximately $3~\mathrm{Mpc}$. Consequently, we can not simply apply the full survey volume, where $V_{\rm max} \propto d_{\rm max}^{3}$. Instead, the dimensionality of the geometric problem reduces from three to one, as the volume along a filament only increases linearly with distance, $V_{\rm max} \propto d_{\rm max}$, rather than with the third power. For a detailed description of the geometry of the filament with respect to our survey volume we refer to the discussion in Appendix~\ref{app_volume}, where approximate analytic expressions for $V_{\rm max}$ as a function of distance are derived.

\begin{figure}
  \includegraphics[width=\linewidth]{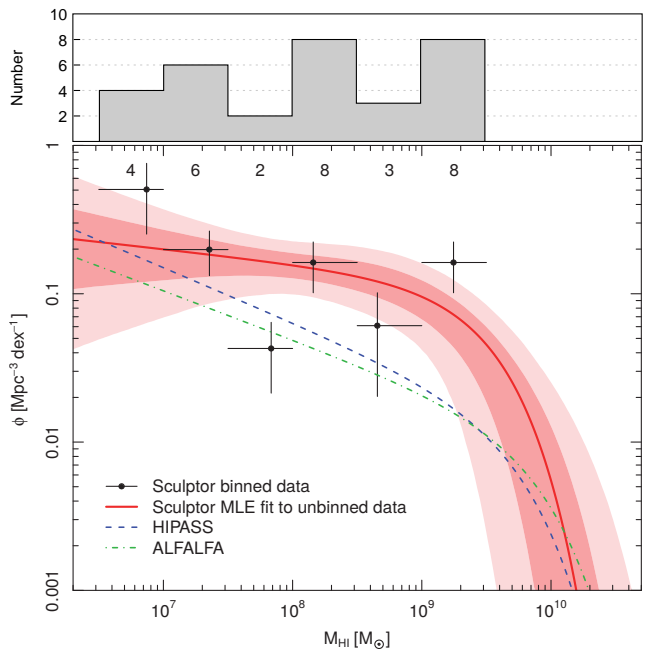}
  \caption{\ion{H}{I} mass function of the Sculptor filament. The black data points show the binned HIMF derived from the catalogued \ion{H}{I} masses, with the numbers and grey histogram denoting the number of galaxies in each mass bin. Horizontal error bars span the bin width of $0.5~\mathrm{dex}$, while vertical error bars correspond to Poisson errors. The red, solid curve shows the Schechter function fitted to the original, unbinned data using a maximum-likelihood estimator, while the shaded regions indicate the 68 and 95~per cent confidence intervals from the fit. The resulting slope is $\alpha = -1.10^{+0.20}_{-0.11}$. The Schechter fits to the HIPASS (blue, dashed curve) and ALFALFA (green, dot-dashed curve) HIMFs from \citet{Zwaan2005} and \citet{Martin2010} are also shown for comparison.}
  \label{fig_himf}
\end{figure}

Assuming $\sigma_{0} = 4.0~\mathrm{mJy}$ and a velocity resolution of $\Delta v_{0} = 26.4~\mathrm{km \, s^{-1}}$ (after Hann filtering) for our HIPASS~2 data, we can then derive the HIMF of the Sculptor group region. The resulting values are presented in Fig.~\ref{fig_himf}, where the black data points show the mean of $\log_{10}(M_{\ion{H}{I}} / \mathrm{M}_{\sun})$ across bins of half a decade, the horizontal error bars span the bin width, and the vertical error bars show the statistical uncertainties from Poisson statistics (with flux and distance errors excluded). Lastly, we fit a Schechter function to the original, unbinned \ion{H}{I} mass measurements from Table~\ref{table_hipass_source_list} using a maximum likelihood estimator (MLE) as described in detail in a separate paper (Obreschkow et al., in prep.). In our case, the log-likelihood function to be maximised is
\begin{equation}
	\ln L(\vec{p}) = \! \int \! \ln \phi(m|\vec{p}) \sum_{i} \varrho_{i}(m|\vec{p}) - \phi(m|\vec{p}) V_{\rm eff}(m) \, \mathrm{d}m
\end{equation}
where $m = \log_{10}(M_{\ion{H}{I}} / \mathrm{M}_{\sun})$ is the normalised, logarithmic \ion{H}{I} mass, $\phi(m|\vec{p})$ is the model to be fitted (in our case a Schechter function with its three free parameters, $\vec{p}$), $\varrho_{i}(m|\vec{p})$ is the probability distribution function of each measurement, $i$, and $V_{\rm eff}(m)$ is the effective maximum volume which we estimate by fitting a continuous function to the individual $1/V_{\rm max}$ measurements. Compared to the standard least-squares minimisation technique there are two great advantages of using a MLE for the fit: (1)~we can account for the statistically significant non-detection of galaxies at the high-mass end of the HIMF, and (2)~the results of the fit are independent of mass binning. The MLE takes into account statistical uncertainties from Poisson statistics, distance errors and flux measurement errors. In addition, we use the fitted Schechter function itself to correct for the effect of Eddington bias on our initial estimate of $\varrho_{i}(m|\vec{p})$ by iteratively repeating the fit until convergence is achieved. The final Schechter function fit resulting from this procedure, plotted as the red curve in Fig.~\ref{fig_himf}, yields the following parameters:
\begin{align}
	\alpha                                                     &= -1.10^{+0.20}_{-0.11} , \\
	\log_{10} \left( M_{\ion{H}{I}}^{\star} \middle/ \mathrm{M}_{\sun} \right) &= +9.53^{+0.22}_{-0.39}, \\
	\log_{10} \left( \Phi^{\star} \middle/ \mathrm{Mpc}^{-3} \right)  &= -1.31^{+0.31}_{-0.22},
\end{align}
where the specified uncertainties correspond to the 16$^{\rm th}$ and 84$^{\rm th}$ percentile. Note that the value of $\Phi^{\star}$, which is normalised per decade of $\log_{10}(M_{\ion{H}{I}} / \mathrm{M}_{\sun})$ here, is somewhat arbitrary, as it depends on our choice of geometry and width of the Sculptor filament. The comparatively small value of $M_{\ion{H}{I}}^{\star}$ simply reflects the sharp drop-off of the observed HIMF beyond $M_{\ion{H}{I}} \approx 3 \times 10^{9}~\mathrm{M}_{\sun}$.

Taking a closer look at the HIMF in Fig.~\ref{fig_himf}, it would appear as if the flatness of the Schechter function was largely driven by a virtually flat HIMF in the intermediate-mass range, whereas near the low-mass end the HIMF of the Sculptor filament appears to be more consistent with the global HIMF from HIPASS and ALFALFA. The density in the highest mass bin in particular is highly significant\footnote{Our survey is volume-limited for galaxies of $M_{\ion{H}{I}} \gtrsim 10^{8}~\mathrm{M}_{\sun}$. In addition, redshift-independent distance measurements are available for all galaxies above $10^{9}~\mathrm{M}_{\sun}$, resulting in small uncertainties in the calculated space density of galaxies.} and largely responsible for shifting the HIMF upwards at the high-mass end, thereby flattening the overall slope of the power-law component of the Schechter function. On the other hand, this effect is reduced again to some degree by the significant non-detection of galaxies with $\log_{10}(M_{\ion{H}{I}} / \mathrm{M}_{\sun}) > 9.5$. Nevertheless, with more than a quarter of all detections in our survey having an \ion{H}{I} mass in excess of $10^{9}~\mathrm{M}_{\sun}$, there are -- on a relative scale -- far too many high-mass galaxies in the Sculptor filament compared to our expectation from the global HIMF.

\section{Discussion}
\label{sect_discussion}

\subsection{Slope of the \ion{H}{I} mass function}
\label{section_himf-slope}

The slope of the HIMF of $\alpha = -1.10^{+0.20}_{-0.11}$, as derived in Section~\ref{sect_himf}, is much flatter than that of the global HIMF from either HIPASS ($\alpha = -1.37 \pm 0.03 \pm 0.05$; \citealt{Zwaan2005}) or ALFALFA ($\alpha = -1.33 \pm 0.02$; \citealt{Martin2010}), both of which are ruled out at the 95~per cent confidence level. The flat slope is consistent with our observation of an almost flat \ion{H}{I} mass distribution along the Sculptor filament with no strong variation in the number density of galaxies across three orders of magnitude in \ion{H}{I} mass (see Fig.~\ref{fig_m-d}). Our result is also consistent with the essentially flat slopes previously found in other group environments \citep{Verheijen2000,Kovac2005,Freeland2009,Kilborn2009,Pisano2011}, thus supporting earlier conclusions that the slope of the HIMF is much flatter in group environments compared to the global galaxy population traced by HIPASS and ALFALFA.

The flat slope of the HIMF implies a lack of \ion{H}{I}-rich low-mass galaxies along the Sculptor filament relative to the observed number of high-mass systems. Several physical mechanisms could potentially explain this effect, including a higher specific star formation rate -- and thus faster consumption of gas -- among low-mass galaxies in low-density environments \citep{Kauffmann2004,Tanaka2004}, or photoionisation of gas in low-mass galaxies following the epoch of reionisation \citep{Benson2002,Kim2013}.

Environmental effects such as ram-pressure stripping or tidal interaction provide another potential mechanism for the removal of gas from galaxies \citep{Gavazzi2005,Hester2006,Fillingham2016}. However, environmentally induced gas removal processes are expected to be more efficient in dense environments and should therefore lead to a flattening of the slope of the HIMF in dense clusters rather than loose groups, e.g.\ as observed in the Virgo Cluster core by \citet{Rosenberg2002}. On the other hand, observational evidence suggests that ram-pressure stripping does affect galaxies in low-density environments \citep{Denes2016} and specifically the outer \ion{H}{I} discs of two of the nearest Sculptor group members, NGC~55 and NGC~300 \citep{Westmeier2011,Westmeier2013}. This result could indicate the presence of an intra-group medium sufficiently dense ($n \approx 10^{-5}~\mathrm{cm}^{-3}$) to possibly have stripped the gas content of low-mass galaxies. The origin of the intra-group medium in the Sculptor group is not known at this stage. The Sculptor filament could potentially constitute a proto-group environment in an early evolutionary stage in which substantial reservoirs of primordial gas still exist in the form of a hot, ionised intra-filament medium that could be responsible for removing or ionising the neutral gas component of galaxies currently falling into the filament.

Curiously, the recent probe into the link between HIMF and environment by \citet{Jones2016}, based on the ALFALFA 70~per cent catalogue, does not find any variation of the slope of the HIMF with environmental density. Their result is in conflict with the studies of individual galaxy groups, including the work presented here. \citet{Jones2016} propose two possible explanations for this discrepancy. On the one hand, their nearest-neighbour method of defining the space density of galaxies could have failed in separating low-density groups from the rest of their sample. On the other hand, the fact that most of the previous group studies were based on interferometric observations could have led to inaccurate detection threshold estimates, as most galaxies would be spatially resolved to a varying degree. However, both our survey as well as the study by \citet{Pisano2011} are based on single-dish observations with the Parkes radio telescope, and most galaxies in our sample are spatially unresolved by our $15.5~\mathrm{arcmin}$ beam (corresponding to a physical resolution in the range of $10$ to $70~\mathrm{kpc}$ along the Sculptor filament), in particular at the crucial low-mass end of the HIMF.

\subsection{Lack of extragalactic \ion{H}{I} gas}

Another notable result of our study is the apparent lack of any neutral gas not associated with galaxies. To begin with, all of our 31~\ion{H}{I} detections have a confirmed or tentative optical counterpart in the Digitized Sky Survey (with the caveat that in five cases no optical redshift measurements are available to unambiguously confirm their association with the \ion{H}{I} source). Hence, we do not detect any so-called `dark galaxies' along the Sculptor filament within the volume covered by our survey.\footnote{In the context of this work, the term `dark' refers to hypothetical dark-matter haloes containing neutral hydrogen, but no detectable stellar component.} Our result is consistent with the general lack of unambiguous detections of such objects in HIPASS \citep{RyanWeber2002,Wong2009} and other \ion{H}{I} surveys \citep{Zwaan2001,Kovac2009}. Potential detections, such as HIPASS~J0731$-$69 \citep{Ryder2001}, are more likely to be the result of tidal or ram-pressure stripping, in particular as they tend to occur near luminous galaxies with obvious signs of recent interaction. More recently, several `dark galaxy' candidates have been reported (e.g., \citealt{Minchin2007,Oosterloo2013}), in particular as part of the ALFALFA survey \citep{Kent2007,Haynes2011,Janowiecki2015}. While the nature of some of these objects is still uncertain, many other `dark galaxy' candidates have since either been identified as tidal debris \citep{Haynes2007,Duc2008,Cannon2015,Taylor2017}, or faint stellar counterparts have been detected in deep optical data \citep{Cannon2015}.

The general lack of observational evidence for `dark galaxies' indicates that such objects either do not exist or are extremely rare. A possible explanation for their rarity could be that low-mass dark-matter haloes are likely to have lost their neutral gas component, either due to tidal or ram-pressure stripping as a result of their shallow gravitational potential \citep{Nickerson2011,BenitezLlambay2013}, or due to photo-ionisation by the UV background radiation field as a result of insufficient shielding \citep{Nickerson2011,BenitezLlambay2017}. Hence, such haloes would not only remain optically dark, but also be undetectable in the \ion{H}{I} 21-cm line.

In addition to the lack of `dark galaxies' in our Sculptor filament data, neither the HIPASS~2 map in Fig.~\ref{fig_smax} nor our combined Parkes and HIPASS~2 map in Fig.~\ref{fig_smax2} reveal any evidence of intergalactic gas in the form of tidal streams or cosmic web filaments. Our deep Parkes observations of the northern Sculptor region have a $5 \, \sigma$ \ion{H}{I} column density sensitivity of about $4 \times 10^{17}~\mathrm{cm}^{-2}$ over $20~\mathrm{km \, s}^{-1}$ for emission filling the $15.5~\mathrm{arcmin}$ beam. This is comparable (within a factor of two) to the sensitivity achieved by \citet{Braun2004} in their deep WSRT survey of the M31/M33 region (approximately $1.5 \times 10^{17}~\mathrm{cm}^{-2}$, assuming $3 \, \sigma$ over $30~\mathrm{km \, s}^{-1}$), but across a larger beam size of $48~\mathrm{arcmin}$.\footnote{Note that their physical beam size at the distance of M31 is comparable to ours at a distance of about $2.5~\mathrm{Mpc}$.} The peak column density along the M31--M33 filament discovered by \citet{Braun2004} is only about $3 \times 10^{17}~\mathrm{cm}^{-2}$, which would be just below our detection limit at the distance of the northern Sculptor group of about $3.5~\mathrm{Mpc}$ and well below our detection limit for more distant parts of the Sculptor filament, in particular as the M31--M33 filament was found to be clumpy rather than diffuse in the GBT follow-up observations presented by \citet{Wolfe2016} at about $9~\mathrm{arcmin}$ angular resolution.

Yet, denser and more massive tidal streams, such as the Magellanic Stream \citep{Mathewson1974}, the gas filaments in the M81 group \citep{Yun1994} or the giant \ion{H}{I} tail in HCG~44 \citep{Serra2013}, would be easily detectable across the entire volume of our survey. Hence, the lack of \ion{H}{I} streams along the entire Sculptor filament within the column density sensitivity and physical beam size of our data suggests that the Sculptor filament constitutes a relatively quiescent environment that has not seen any recent major mergers or tidal interaction between its more massive members. This finding could lend support to the proto-group scenario proposed in Section~\ref{section_himf-slope} and would not contradict previous observational evidence of small-scale mergers and interactions having affected individual galaxies in the Sculptor group, including NGC~55 \citep{Tanaka2011,Westmeier2013}, NGC~253 \citep{Prada1998} and NGC~300 \citep{Rogstad1979,Westmeier2011}. Note that the situation changes drastically when we follow the filament in the direction opposite to the Sculptor group, where we encounter several more advanced and strongly interacting systems, including M31/M33 \citep{Braun2004,Lewis2013}, M51/NGC~5195 \citep{Haynes1978,Rots1990}, M81/M82 \citep{Yun1994} and, first and foremost, our own Milky Way which is currently interacting with the Large and Small Magellanic Clouds \citep{Mathewson1974}.

\section{Summary and conclusions}
\label{sect_summary}

In this paper we present the results of a deep \ion{H}{I} study of the nearby Sculptor group and filament based on a deep survey of the northern Sculptor group region with the Parkes telescope and an improved, more sensitive version of the \ion{H}{I} Parkes All-Sky Survey \citet{Barnes2001} referred to as HIPASS~2. Our main results and conclusions are summarised below.
\begin{enumerate}
  \item Overall, we detect 31~\ion{H}{I} sources within the Sculptor filament, eight of which are new \ion{H}{I} detections. The \ion{H}{I} masses of the sources range from about $4 \times 10^{6}$ to $3 \times 10^{9}~\mathrm{M}_{\sun}$, and their distances are in the range of about $2$ to $15~\mathrm{Mpc}$ along a cosmic filament roughly parallel to the line of sight that we refer to as the Sculptor filament.
  \item We derive the HIMF of the Sculptor filament under the assumption that the galaxies are located in a one-dimensional filament with a thickness of $3~\mathrm{Mpc}$ that is oriented along the line of sight. We find that the slope of the HIMF of $\alpha = -1.10^{+0.20}_{-0.11}$ is much flatter than for the global galaxy samples traced by large surveys such as HIPASS \citep{Zwaan2005} and ALFALFA \citep{Martin2010}, and consistent with the flat slopes found in low-density group environments by other studies \citep{Verheijen2000,Kovac2005,Kilborn2009,Freeland2009,Pisano2011}.
  \item Several mechanisms could potentially explain this flattening of the slope of the HIMF in group environments, including higher efficiency of star formation in low-mass group galaxies, or photoionisation of the gas due to insufficient shielding of low-mass satellites in low-mass parent haloes. Lastly, the Sculptor filament could constitute a proto-group environment with substantial reservoirs of hot intra-filament gas that could strip or ionise the neutral gas component of low-mass galaxies falling into the filament. Our result of a flat HIMF slope is in tension with a recent study by \citet{Jones2016} based on ALFALFA data, which finds no variation in the slope of the HIMF with environmental density.
  \item After cross-matching our source catalogue with optical galaxy catalogues, we find that all detected sources have a potential optical counterpart at the right location and redshift, although in five cases there is no optical redshift information available and an association is assumed based on spatial coincidence alone. Hence, all \ion{H}{I} detections are likely associated with galaxies. The apparent lack of so-called `dark galaxies' in the Sculptor filament is in good agreement with similar results from previous \ion{H}{I} studies of other galaxy samples. \citep{Zwaan2001,RyanWeber2002,Wong2009,Kovac2009}. Nevertheless, dark-matter haloes without stars might still exist, but be undetectable in \ion{H}{I} emission as the result of ionisation or stripping of their neutral gas.
  \item Despite a superb $5 \, \sigma$ column density sensitivity of about $4 \times 10^{17}~\mathrm{cm}^{-2}$ towards the northern part of the Sculptor group, we do not detect any intergalactic gas or tidal streams, suggesting that the Sculptor filament constitutes a relatively quiescent environment that has not seen any recent major mergers or accretion events, although individual galaxies may still have accreted satellites on a smaller scale. Faint structures such as the M31--M33 filament discovered by \citet{Braun2004} would likely remain undetectable in our data, but more massive structures such as the Magellanic Stream would have been detected.
\end{enumerate}
In combination with new multi-wavelength data, the upcoming wide and/or deep \ion{H}{I} surveys on precursor and pathfinder facilities of the Square Kilometre Array will soon enable us to study the environmental effects that shape the neutral gas content and star formation rate of galaxies in far greater detail and across a large range of different environments and redshifts. Combining unprecedented angular resolution and sensitivity, the new data will finally allow the issue of how the HIMF changes across environments of varying density to be settled and help us to pin down the physical processes that are ultimately responsible for the removal and processing of the neutral gas component in galaxies.

\section*{Acknowledgements}

The Parkes radio telescope and the Australia Telescope Compact Array are part of the Australia Telescope National Facility which is funded by the Australian Government for operation as a National Facility managed by CSIRO.

Based on photographic data obtained using The UK Schmidt Telescope. The UK Schmidt Telescope was operated by the Royal Observatory Edinburgh, with funding from the UK Science and Engineering Research Council, until 1988 June, and thereafter by the Anglo-Australian Observatory. Original plate material is \textcopyright{}~the Royal Observatory Edinburgh and the Anglo-Australian Observatory. The plates were processed into the present compressed digital form with their permission. The Digitized Sky Survey was produced at the Space Telescope Science Institute under US Government grant NAG~W--2166.

This research has made use of the NASA/IPAC Extragalactic Database (NED) which is operated by the Jet Propulsion Laboratory, California Institute of Technology, under contract with the National Aeronautics and Space Administration.

\appendix

\section[]{Calculation of the maximum detection volume}
\label{app_volume}

\begin{figure*}
  \includegraphics[width=0.75\linewidth]{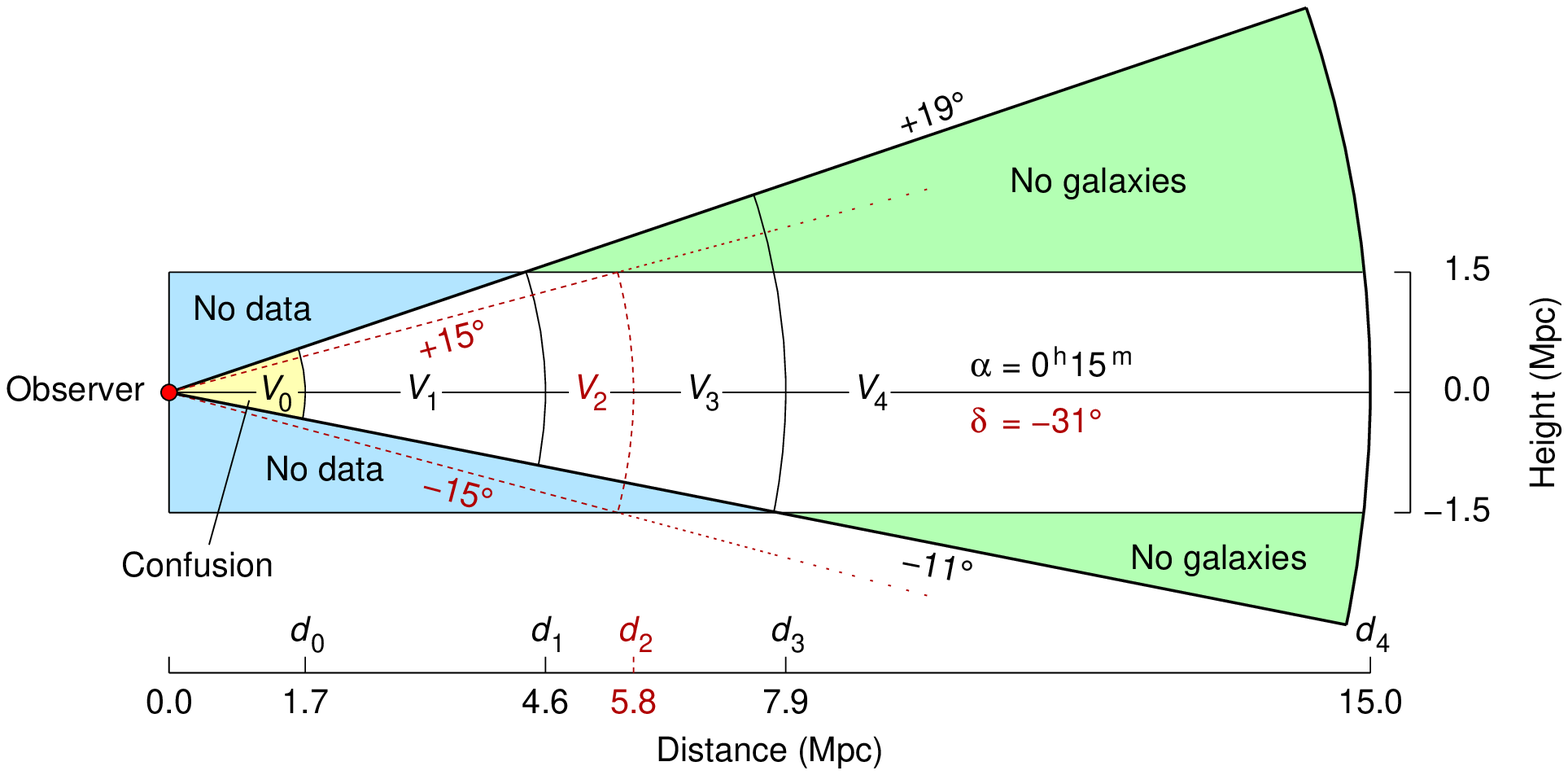}
  \caption{Geometry of our survey volume and the embedded Sculptor filament (see Fig.~\ref{fig_filament} for comparison). The situation shown here is the projection in right ascension, where we face a slight offset between the centre of our survey area and the centre of the filament of $\Delta \alpha \approx 0^{\rm h}15^{\rm m}$. In declination our survey area is roughly centred on the filament at $\delta \approx -31\degr$, as outlined by the dashed, red lines in the figure. The white area corresponds to the overlap region between the filament and our survey volume.}
  \label{fig_geometry}
\end{figure*}

In this section we calculate the maximum volume, $V_{\rm max}$, along the Sculptor filament corresponding to a specific maximum distance, $d_{\rm max}$, taking into account the geometry of both the filament and our survey volume as illustrated in Fig.~\ref{fig_geometry}. The filament is assumed to be oriented along the line of sight away from the observer. We further assume that the filament has the shape of a cuboid with a width and height of $3~\mathrm{Mpc}$ and a length of $15~\mathrm{Mpc}$ (corresponding to the distance of the far end of the filament from the Milky Way). In addition, we make the assumption that our survey area in HIPASS~2 covers about $30\degr \times 30\degr$ on the sky, thereby cutting out a volume of space corresponding to that of a rectangular spherical pyramid with an opening angle of $30\degr$ and a height equal to $15~\mathrm{Mpc}$ (discarding any volume beyond the far end of the filament).

\begin{figure}
  \includegraphics[width=\linewidth]{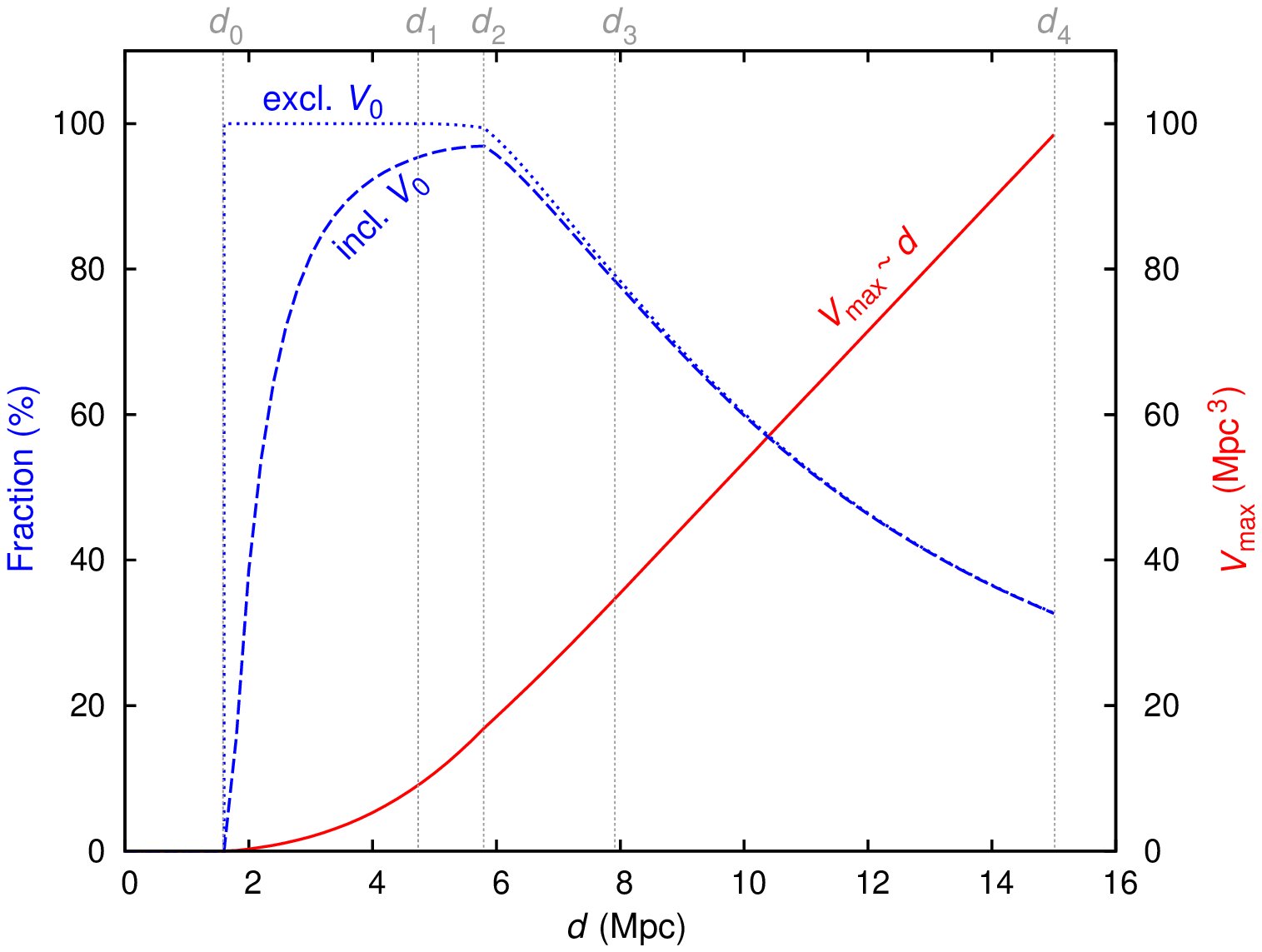}
  \caption{Maximum detectable volume, $V_{\rm max}$, of the Sculptor filament as a function of distance, $d$ (red, solid line). The fraction of $V_{\rm max}$ over our total survey volume is shown as the blue curves for two cases: including (dashed, blue line) and excluding (dotted, blue line) the confusion-affected volume, $V_{0}$ (compare Fig.~\ref{fig_geometry}). Note that beyond $d_{3}$ the value of $V_{\rm max}$ increases only linearly with distance, as the filament is fully embedded in our survey volume.}
  \label{fig_vmax}
\end{figure}

The region in space where the Sculptor filament and our survey volume overlap will be considered as the detectable volume of the filament (shown as the white region in the two-dimensional projected view in Fig.~\ref{fig_geometry}). There are three different regions where we do not detect any galaxies:
\begin{enumerate}
  \item Parts of the near end of the filament are so close that they extend beyond the edges of our survey area. Hence, our coverage of the filament becomes incomplete below a certain distance. This region is highlighted in blue in Fig.~\ref{fig_geometry}.
  \item At larger distances our survey volume extends well beyond the boundaries of the filament, and we therefore do not detect any galaxies across significant parts of our survey volume that are outside the filament. This region is highlighted in green in Fig.~\ref{fig_geometry}.
  \item Below a Hubble distance of about $1.7~\mathrm{Mpc}$ we suffer from confusion with foreground emission from the Milky Way and the Magellanic Stream, preventing us from detecting galaxies below that distance limit. This region is highlighted in yellow in Fig.~\ref{fig_geometry}.
\end{enumerate}
Our aim therefore is to calculate for each possible distance the white detectable volume that lies within that distance from the Milky Way. For this purpose we can split up that volume into sub-volumes that can either be described by a spherical pyramid with opening angles $\phi_{\alpha}$ and $\phi_{\delta}$ and height $h$ (describing the geometry of our survey volume), or approximated by a Cartesian cuboid with width $w$, height $h$ and length $l$ (describing the geometry of the Sculptor filament). Note that we assume a cuboid rather than a cylinder merely for the sake of simplicity, and this choice is not expected to have any significant impact on the derived slope of the HIMF.\footnote{It will only affect the global normalisation factor which is by definition arbitrary, as it depends on the assumed geometry of the filament relative to our survey volume.} The volume of a spherical pyramid is given by
\begin{align}
  V_{\rm pyr} & = \frac{4}{3} \, h^{3} \arcsin \! \left[ \sin \! \left( \frac{\phi_{\alpha}}{2} \right) \sin \! \left( \frac{\phi_{\delta}}{2} \right) \right] \nonumber \\
              & \approx \frac{4}{3} \, h^{3} \sin \! \left( \frac{\phi_{\alpha}}{2} \right) \sin \! \left( \frac{\phi_{\delta}}{2} \right) , \label{eqn_pyramid}
\end{align}
where we made use of the small-angle approximation of the sine function, while the volume of a Cartesian cuboid is simply defined as $V_{\rm cub} = w \times h \times l$. From this we can now specify the volumes corresponding to different distances along the line of sight. According to Eq.~\ref{eqn_pyramid}, the volume of the yellow, confusion-affected region below $1.7~\mathrm{Mpc}$ distance is given by
\begin{equation}
  V_{0} \approx \frac{4}{3} \, d_{0}^{3} \sin^{2}(15\degr) \approx 0.44~\mathrm{Mpc}^{3}
\end{equation}
and will need to be subtracted from any volume calculation, as we can not detect galaxies within that region. Subsequent volumes, as defined in Fig.~\ref{fig_geometry}, are given by
\begin{align}
  V_{1} \approx & \; \frac{4}{3} \, d_{1}^{3} \sin^{2}(15\degr) , \\
  V_{2} \approx & \; \frac{4}{3} \sin(15\degr) \! \left[ d_{2}^{3} \sin(15\degr) - (d_{2} - d_{1})^{3} \sin(9.5\degr) \right] \! , \\
  V_{3} \approx & \; V_{2} + \frac{D}{2} \left[ D (d_{3} - d_{2}) + \sin(11\degr) \left( d_{3}^{2} - d_{2}^{2} \right) \right] \! , \\
  V_{4} \approx & \; V_{3} + D^{2} (d_{4} - d_{3}) ,
\end{align}
where $D = 3~\mathrm{Mpc}$ is assumed to be the thickness of the Sculptor filament. Note that all volumes are defined to span the entire distance range from the observer to their respective outer boundary and, as mentioned before, the value of $V_{0}$ will still need to be subtracted to exclude the nearby volume affected by Galactic and Magellanic confusion. As shown in Fig.~\ref{fig_vmax}, the detectable volume of the filament enclosed within the effective volume covered by our survey amounts to approximately $100~\mathrm{Mpc}^{3}$, which corresponds to about one third of our effective survey volume of $300~\mathrm{Mpc}^{3}$ within the assumed distance limit of $15~\mathrm{Mpc}$.

\section[]{Optical and \ion{H}{I} maps}
\label{app_maps}

For all sources detected in our survey of the Sculptor filament we show in Fig.~\ref{fig_dss-images} an optical image from the Digitized Sky Survey with the \ion{H}{I}~contours from our HIPASS~2 data overlaid.

\begin{figure*}
  \includegraphics[width=\linewidth]{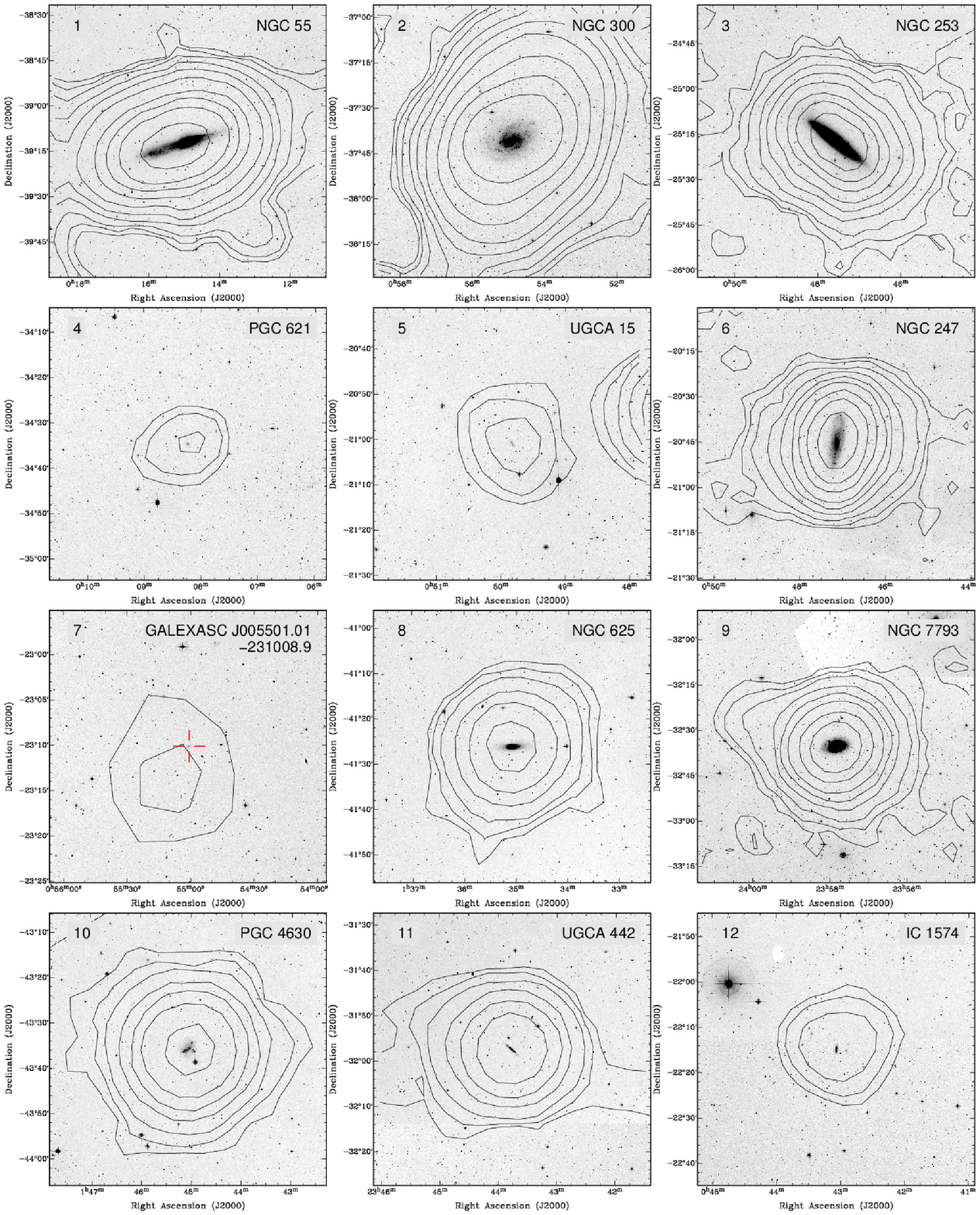}
  \caption{Optical images from the Digitized Sky Survey (DSS) of all potential optical counterparts, with the \ion{H}{I} column density contours of the corresponding HIPASS~2 detections overlaid. The lowest contour level is $5 \times 10^{16}~\mathrm{cm}^{-2}$, and subsequent contour levels are drawn at $1$, $2$ and $5 \times 10^{n}~\mathrm{cm}^{-2}$, where $n \in \mathbb{N}$, $n \ge 17$. In a few cases, fainter or more uncertain counterparts are marked with red cross-hairs. Note that the size of the maps varies ($0.5\degr{}$, $1.0\degr{}$ or $1.5\degr{}$) depending on the size and brightness of the depicted galaxy.}
  \ContinuedFloat
  \label{fig_dss-images}
\end{figure*}

\begin{figure*}
  \includegraphics[width=\linewidth]{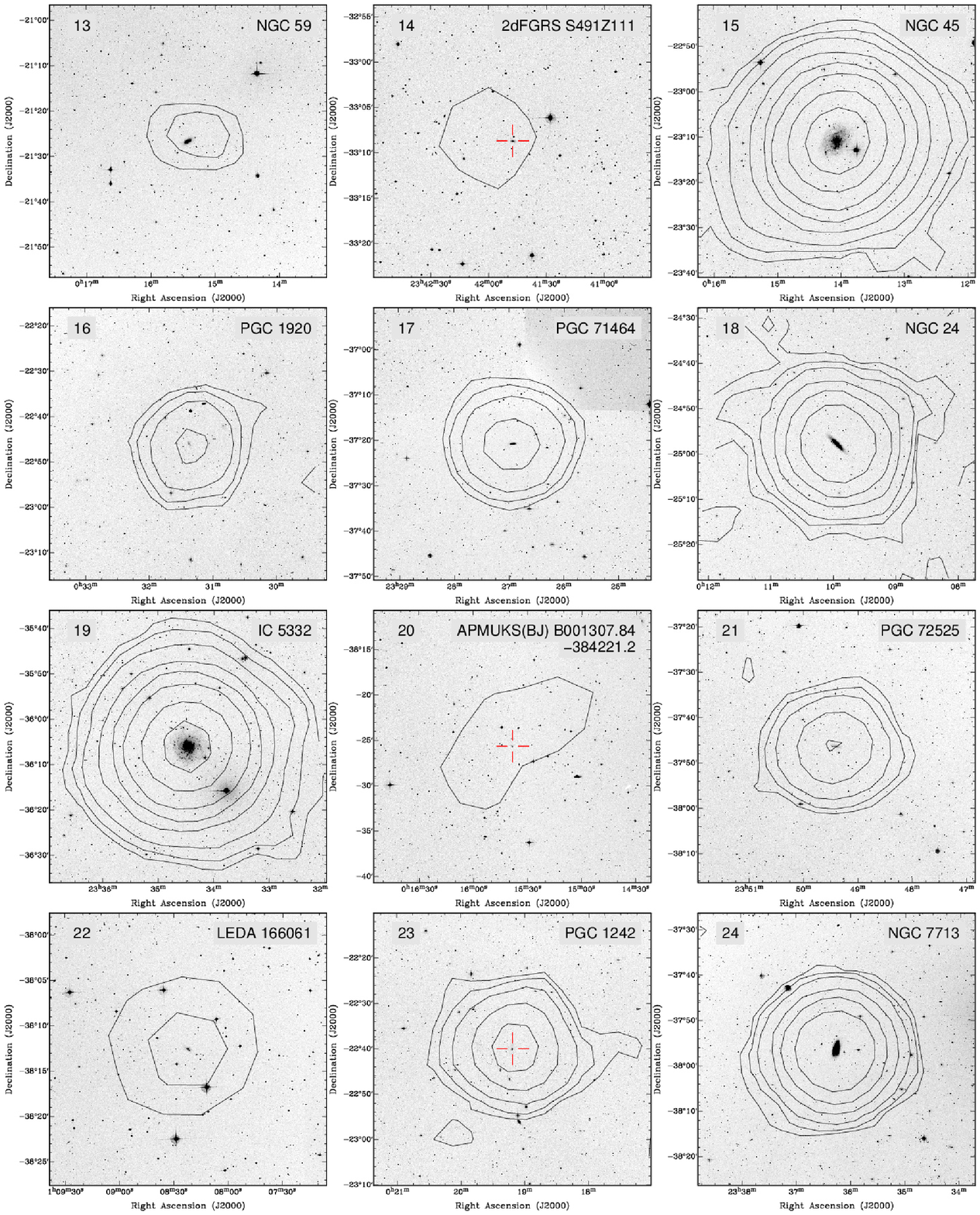}
  \caption{(continued)}
  \ContinuedFloat
\end{figure*}

\begin{figure*}
  \includegraphics[width=\linewidth]{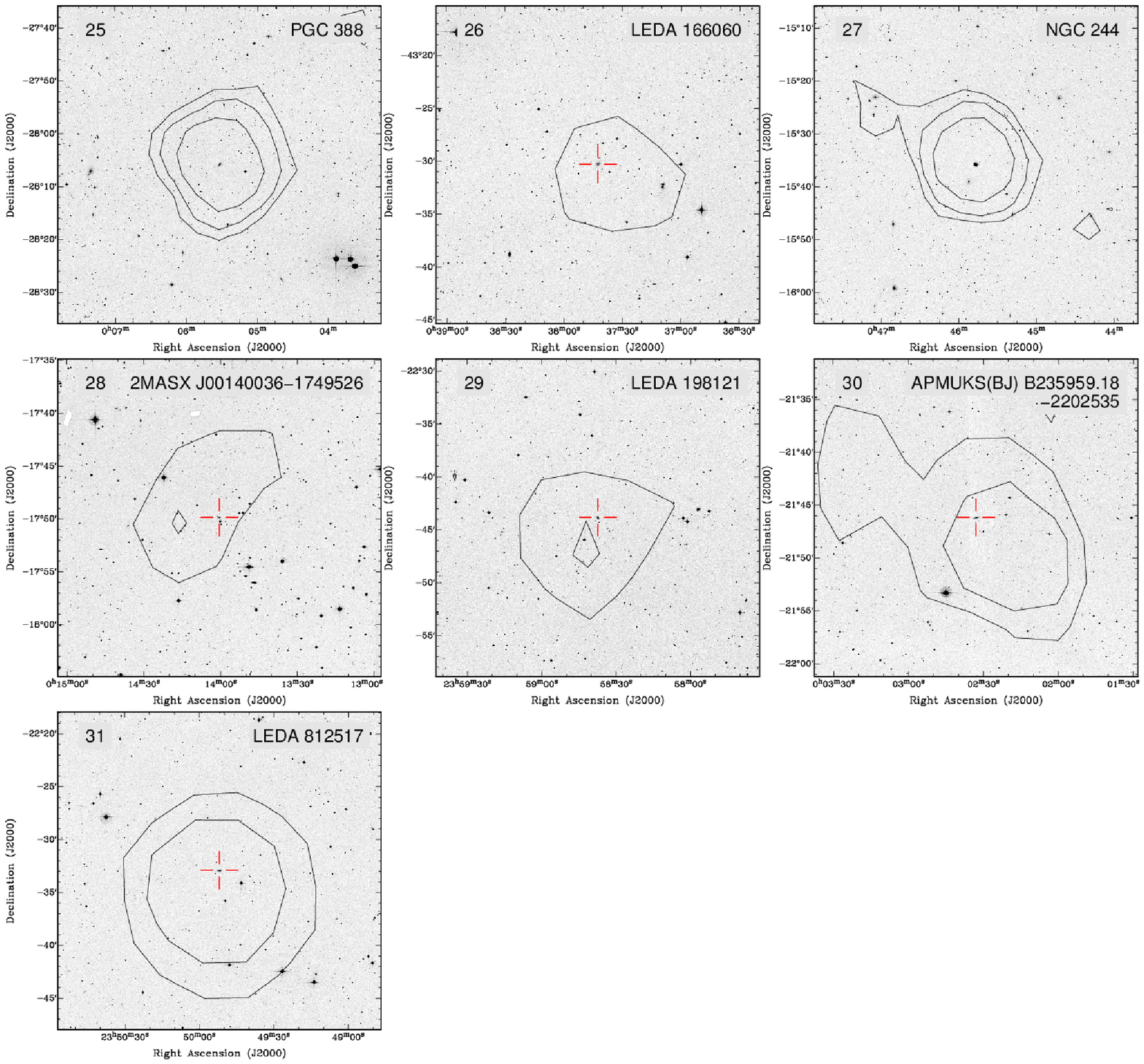}
  \caption{(continued)}
  \ContinuedFloat
\end{figure*}

\section[]{Integrated spectra}
\label{app_spectra}

Integrated HIPASS~2 \ion{H}{I}~spectra of all sources detected in our survey of the Sculptor filament are presented in Fig.~\ref{fig_hipass_spectra}. Each spectrum was fitted with a Busy Function \citep{Westmeier2014} for the purpose of parameterisation.

\begin{figure*}
  \includegraphics[width=0.9\linewidth]{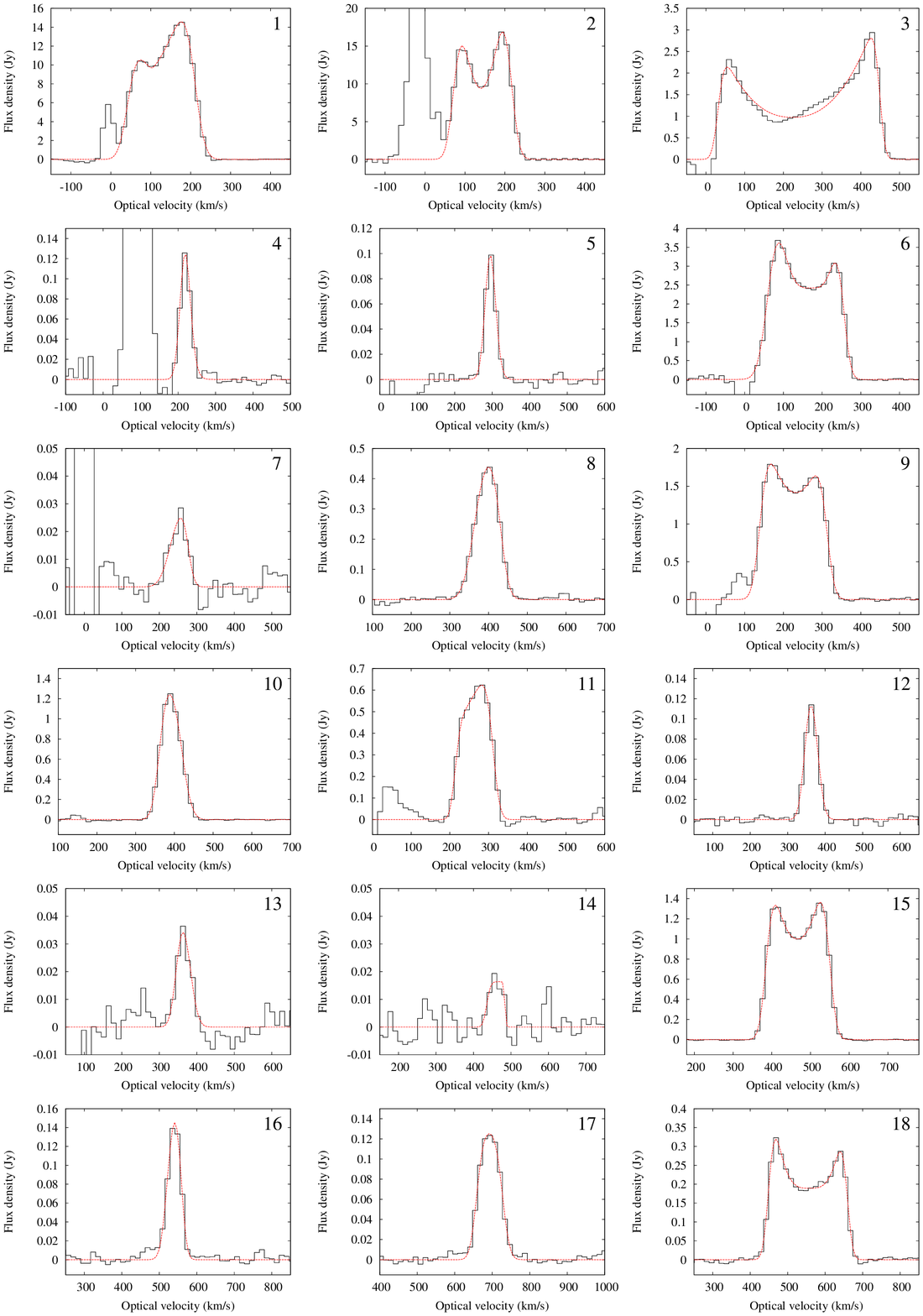}
  \caption{Integrated \ion{H}{I} spectra of all galaxies found the HIPASS~2 data of the Sculptor group region. The red, dashed line shows the Busy Function \citep{Westmeier2014} fitted to the data for the purpose of parameterisation. Strong positive and negative signals near velocities of $\mathrm{c}z \approx 0~\mathrm{km \, s}^{-1}$ are due to Galactic or Magellanic foreground emission.}
  \ContinuedFloat
  \label{fig_hipass_spectra}
\end{figure*}

\begin{figure*}
  \includegraphics[width=0.9\linewidth]{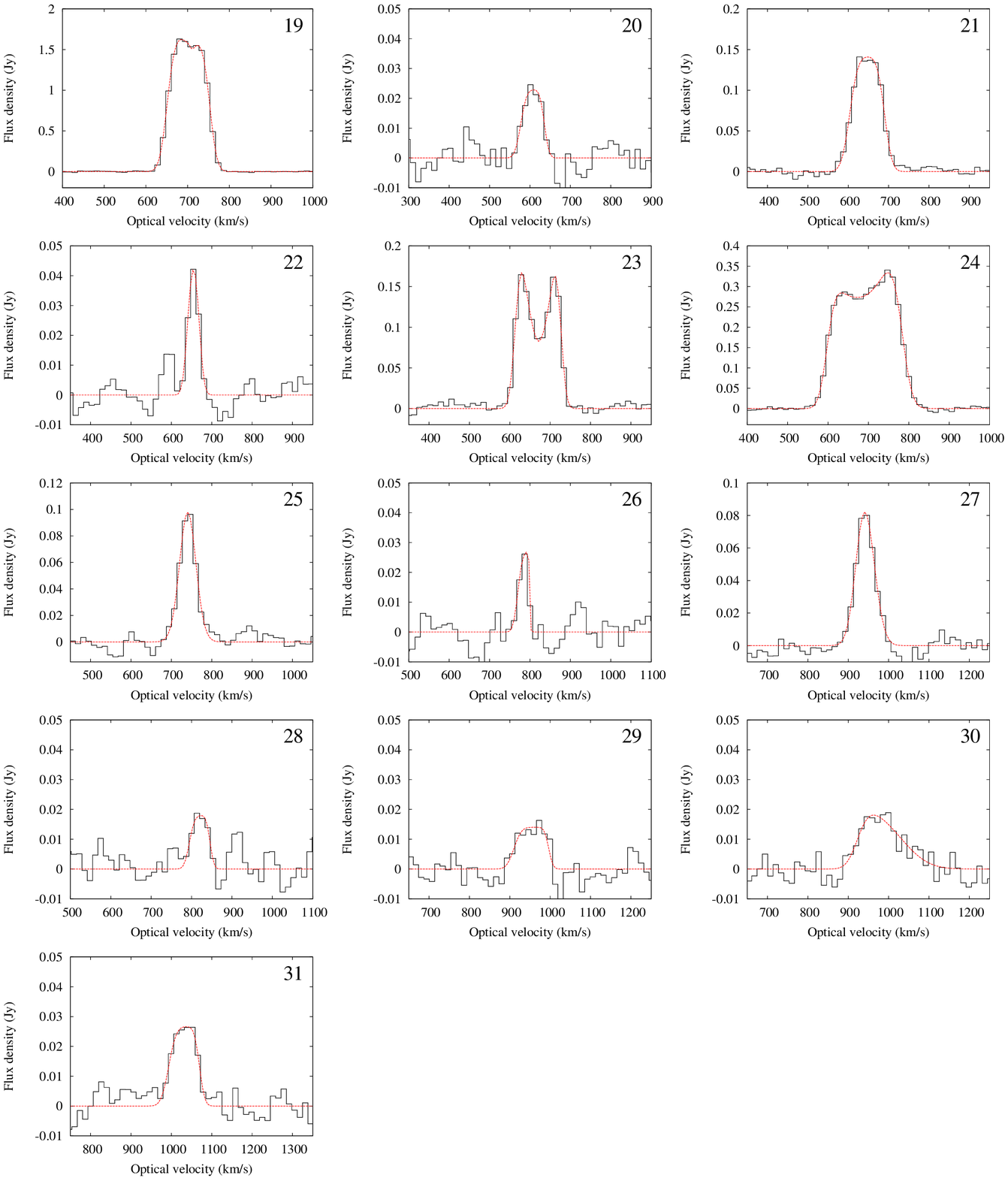}
  \caption{(continued)}
  \ContinuedFloat
\end{figure*}

\bsp

\label{lastpage}

\end{document}